\documentclass[%
reprint,
superscriptaddress,
 amsmath,amssymb,
 aps,prl
]{revtex4-1}

\usepackage{mathtools}
\usepackage{graphicx}
\usepackage{dcolumn}
\usepackage{bm}

\usepackage{amsmath}
\usepackage{amssymb}
\usepackage{amsthm}
\usepackage{graphicx}
\usepackage{algorithmic}
\usepackage{mathrsfs}
\usepackage{braket}
\usepackage{xcolor}
\usepackage{comment}
\usepackage{mathtools, nccmath}
\usepackage{tikz}
\usepackage[roman]{complexity}
\usepackage[caption=false]{subfig}
\usepackage{physics}

\usepackage{pdfpages}
\usepackage{pgffor}
\makeatletter
\AtBeginDocument{\let\LS@rot\@undefined}
\makeatother

\begin{document}

\title{Quantum Kerr Learning}

\author{Junyu Liu}
\email{junyuliu@uchicago.edu}
\affiliation{Pritzker School of Molecular Engineering, The University of Chicago, Chicago, IL 60637, USA}
\affiliation{Chicago Quantum Exchange, Chicago, IL 60637, USA}
\affiliation{Kadanoff Center for Theoretical Physics, The University of Chicago, Chicago, IL 60637, USA}
\affiliation{qBraid Co., Harper Court 5235, Chicago, IL 60615, USA}

\author{Changchun Zhong}
\affiliation{Pritzker School of Molecular Engineering, The University of Chicago, Chicago, IL 60637, USA}

\author{Matthew Otten}
\affiliation{HRL Laboratories, LLC, Malibu, CA 90265, USA}

\author{Anirban Chandra}
\affiliation{Center for Nanoscale Materials, Argonne National Laboratory, Lemont, Illinois 60439, USA}
\affiliation{University of Illinois, Chicago, Illinois 60607, USA}

\author{Cristian L. Cortes}
\affiliation{Center for Nanoscale Materials, Argonne National Laboratory, Lemont, Illinois 60439, USA}

\author{Chaoyang Ti}
\affiliation{Center for Nanoscale Materials, Argonne National Laboratory, Lemont, Illinois 60439, USA}

\author{Stephen K Gray}
\affiliation{Center for Nanoscale Materials, Argonne National Laboratory, Lemont, Illinois 60439, USA}


\author{Xu Han}
\email{xu.han@anl.gov}
\affiliation{Center for Nanoscale Materials, Argonne National Laboratory, Lemont, Illinois 60439, USA}

\date{\today}

\begin{abstract}

Quantum machine learning is a rapidly evolving field of research that could facilitate important applications for quantum computing and also significantly impact data-driven sciences. In our work, based on various arguments from complexity theory and physics, we demonstrate that a single Kerr mode can provide some ``quantum enhancements'' when dealing with kernel-based methods. Using kernel properties, neural tangent kernel theory, first-order perturbation theory of the Kerr non-linearity, and non-perturbative numerical simulations, we show that quantum enhancements could happen in terms of convergence time and generalization error. Furthermore, we make explicit indications on how higher-dimensional input data could be considered. Finally, we propose an experimental protocol, that we call \emph{quantum Kerr learning}, based on circuit QED.

\end{abstract}

\maketitle


\emph{Introduction.---}Quantum machine learning, i.e., combining machine learning with the computational power of quantum devices, 
is an exciting and emerging direction for modern information technology~\cite{huang2020power,abbas2021power,liu2021rigorous,aharonov2022quantum}. In the present Noisy Intermediate-Scale Quantum (NISQ) era \cite{preskill2018quantum}, there has been significant progress on quantum machine learning based on variational quantum circuits \cite{schuld2015introduction,biamonte2017quantum,dunjko2018machine,farhi2018classification,cong2019quantum,otten2020quantum,bausch2020recurrent,beer2020training,Mangini_2021,Liu:2021wqr}. However, it is still not completely clear, both in theory and practice, if and how a true quantum advantage relative to completely classical-computer based approaches will be achieved \cite{mcclean2018barren,Liu:2021wqr}.

Over several years in classical machine learning, kernel methods have been developed and applied to address numerous problems \cite{mohri2018foundations}. A non-linear machine learning problem could ideally be described by linear models in a sufficiently high dimensional \emph{Hilbert space} using these kernel methods. In the context of classical machine learning, the corresponding Hilbert space of a given kernel is abstract, whereas in the quantum computing setup, the Hilbert space could be \emph{physical} -- such as the state vector space of a quantum device \cite{havlivcek2019supervised}.    Complex quantum systems could produce complicated enough kernels that are hard to evaluate using classical computers, providing a potential regime of quantum advantage in quantum machine learning \cite{liu2021rigorous}. 

Now, how can we produce complicated kernels in a quantum system? In the quantum kernel method, the kernel is evaluated from a quantum measurement. For instance, if we consider a Hamiltonian with the real time evolution, $H(t)$, one could construct the kernel as

\begin{equation}
\begin{split}
    K \left( \mathbf{x}, \mathbf{x}' \right) =\, & \bigg| \bigg\langle \psi \bigg|
    \bar{\mathbb{T}}\exp \left[ {\frac{i}{\hbar }\textstyle{\int_0^{T'} {H(t')} dt'}} \right] \\
    & \times 
    \mathbb{T}\exp \left[  - \frac{i}{\hbar }\textstyle{\int_0^T {H(t)} dt} \right]\bigg|\psi
    \bigg\rangle\bigg|^2.
    \label{eq:Kernel}
\end{split}
\end{equation}
Here, starting from $\ket{\psi}$ and evolving with the Hamiltonian $H(t)$ during over times $T$ and $T'$ respectively, the inner product of two quantum states is measured. $\mathbb{T}$ denotes time ordering in quantum mechanics. The more complicated the Hamiltonian is, the harder it is for a classical computer to simulate. In Equation \ref{eq:Kernel}, the kernel $K$ is a matrix where the matrix elements are specified by the input data points, $\mathbf{x}$ and $\mathbf{x}'$. The vectors $\mathbf{x}$ and $\mathbf{x}'$ could be constructed from parameters of the Hamiltonian or the evolution time. 

\textit{Model setup.---}In this paper, we consider a single-mode quantum system with Kerr non-linearity that could potentially provide computational benefits compared to its classical counterparts. Such systems can be easily realized in circuit QED experiments where the Kerr non-linearity can be obtained from Josephson junctions or the kinetic inductance of a superconducting resonator \cite{Xu2019,Han_2020,Han2022}. We propose and describe an experimental realization in the supplemental material (SM). The effective Hamiltonian of the considered system can be described as,
\begin{align}\label{hamiltonian}
&H = {H_\mathrm{f}} + {H_\mathrm{I}}~, {H_\mathrm{f}} = {H_0} + {H_\mathrm{t}}\nonumber\\
&\frac{{{H_0}}}{\hbar } = {\omega _\mathrm{m}}\left( {{b^\dag }b + \frac{1}{2}} \right)~,\frac{{{H_\mathrm{t}}}}{\hbar } = \Omega \left( {b{e^{  i{\omega _\mathrm{L}}t}} + {b^\dag }{e^{-i{\omega _\mathrm{L}}t}}} \right)~,\nonumber\\
&\frac{{{H_\mathrm{I}}}}{\hbar } = {-K_{{\rm{err}}}}{b^\dag }{b^\dag }bb~.
\end{align}
For our demonstrative examples, we restrict ourselves to  three-dimensional inputs and treat $\omega_\mathrm{m}$ and $K_{\text{err}}$ as constants. The input data is thus defined as a real vector $\mathbf{x}=(\mathbf{x})_{i=1}^3=(\Omega ,{\omega _\mathrm{L}},T)$. $\Omega$ and $\omega_{\mathrm{L}}$ standard for the energy and the frequency of the laser, and $T$ means the evolution time. All of them are tunable in the experimental devices.

Depending on the magnitude of $K_{\text{err}}$ a few regimes exist that warrant some discussion. When the Kerr non-linearity is turned off, the system Hamiltonian reduces to a quadratic form in the rotating frame. If we turn on the Kerr non-linearity and when $K_{\text{err}}$ is much smaller than $\Omega$, the kernel could be predicted by perturbation theory (see SM for detailed calculations). On the other hand, when $K_{\mathrm{err}}$ is much larger than $\Omega$, the non-linear resonator can be approximated as a qubit with negligible excitations at energy levels higher than the first excited state (see SM). However, for a general $K_{\text{err}}$ in the regime of $K_{\text{err}}/ \Omega 	\sim \mathcal{O}(1)$, we do not have analytic predictions and the system must be evaluated on a quantum system or simulated using a classical computer.

\emph{Quantum Enhancement.---}  The Kerr non-linear quantum system considered in this work has some inherent advantages that imply quantum enhancement. Here, we use the word ``quantum enhancement" to indicate benefits in computation provided by the Kerr non-linearity; this is not the same as ``quantum advantage" determined by the rigorous complexity theory statements and practical demonstrations. However, a few complexity-related arguments are presented in SM. A few possible inherent advantages of our Kerr system are discussed next. 


First, although in the bulk of our paper, we treat $\Omega$ as a constant, we could make it time-dependent. For the time-dependent $\Omega (t)$, it is shown that the Hamiltonian in Eq.\,(\ref{hamiltonian}) could simulate universal quantum computing in polynomial time, and the whole class \textbf{BQP} \cite{lloyd1999quantum}, even for small Kerr couplings. Thus, we expect that a combination of time-dependent $\Omega(t)$ and $K_{\text{err}}$ could provide further quantum enhancements.

Second, although the bulk of our paper is about single modes, we briefly introduce and discuss multimodes. A multimode Hamiltonian in our circuit QED setup will include several free bosonic modes that are interacting with another Kerr mode. Through the coupling, those free modes will have more complicated dynamics, and we expect that it might present stronger quantum enhancements. An analog of the multiple mode system will be a zero-dimensional $\lambda \phi^4$ quantum field theory on the lattice with the system size $N$ (note that the Kerr term is also quartic), where we have the theoretical predictions when the bare coupling $\lambda$ is either zero or infinity. The non-perturbative corrections will happen around the critical point, and the strong coupling regime, where we have a $\mathbb{Z}_2$ symmetry which is spontaneously broken, and we expect a quantum speedup to simulate the non-perturbative dynamics \cite{Jordan:2012xnu,Jordan:2011ci,Preskill:2018fag}.

\emph{Kernel Properties and Statistics.---} The property of the kernel function $K$, or more precisely the neural tangent kernel (NTK) $K_\mathrm{H}=K^2$ of the linear model (see SM), plays an important role in the performance of the machine learning algorithm. To discuss and demonstrate the quantum enhancement achieved by Kerr non-linearity, we first create an arbitrary kernel matrix by specifying the associated parameters -- ~$\Omega, ~\omega _\mathrm{L}, ~T, ~\omega_\mathrm{m}, ~K_{\text{err}}$. We randomly generate the input data $\mathbf{x}=\left(\Omega, \omega_{\mathrm{L}}, T\right)$ through uniform distributions within experimentally feasible data range. We set the data range from $(0,0,0)$ to $\left(\Omega^{\text{range}}, \omega_{\mathrm{L}}^{\text{range}}, T^{\text{range}}\right)\sim (300\, \text{MHz}\times2\pi,10\, \text{GHz}\times2\pi, 0.05\,\mu \text{s})$. Moreover, we fix $\omega_\mathrm{m}$ at around $10\, \text{GHz}\times2\pi$ which is typical for a superconducting microwave resonator. $K_{\text{err}}$ is treated as a control parameter to understand potential quantum enhancement.

\begin{figure}[t]
\centering
\includegraphics[width=0.48\textwidth]{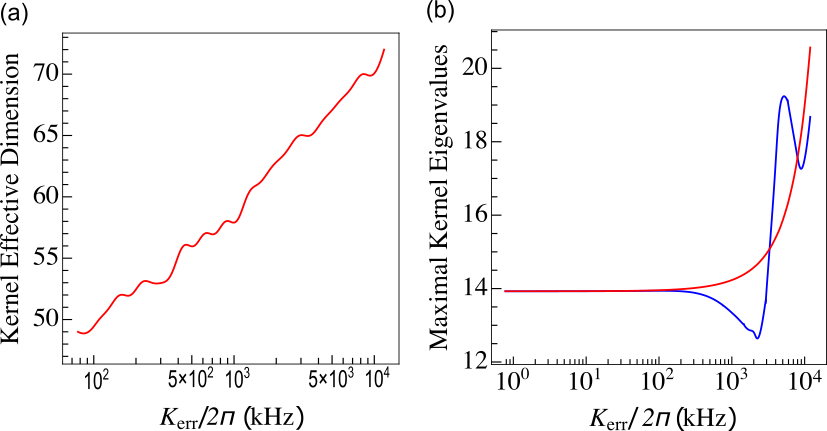}
\caption{Kernel statistics when turning on Kerr non-linearity. We generate $\mathcal{O}(100)$ data vectors when evaluating the kernel statistics using the perturbation theory (red), and $\mathcal{O}(10)$ data vectors when evaluating the kernel statistics using the numerical exact simulation with 100 truncated bosonic energy levels (blue). As evidence of where nontrivial Kerr coupling complexifies the kernel, we (a) count the number of eigenvalues of $K_\mathrm{H}$ that are larger than $10^{-7}$ (the kernel effective dimension) when turning on the Kerr coupling, and (b) evaluate the maximal eigenvalues of $K_\mathrm{H}$ depending on $K_{\text{err}}$. Note that when $K_{\text{err}} \ge \mathcal{O}(0.1)\, \text{MHz}\times 2\pi$, we start entering the regime of non-perturbative dynamics where perturbative theory predictions are not trustworthy.}
\label{fig:kerstat}
\end{figure}

We assess the performance of our proposed $K_{\text{err}}$ kernel based on the theory of the kernel method \cite{mohri2018foundations} and the neural tangent kernel theory for linear models \cite{lee2017deep,jacot2018neural,lee2019wide,arora2019exact,sohl2020infinite,yang2020feature,yaida2020non,roberts2021principles,Liu:2021wqr}. Note that, we use the traditional quantum kernel methods to solve the problem that is similar to the support vector machine with continuous variables, but we use neural tangent kernel theory as a tool to make predictions about the gradient descent dynamics in linear regression. Neural tangent kernel theory is a theory for explaining wide neural networks, which could be deep but the ratio of depth to width should be small. In the large-width limit, the gradient descent dynamics of the neural networks will coincide with the kernel method, so it is also called the kernel limit. In our work, we directly use a linear model (namely, the kernel method), so the square of the neural tangent kernel is equal to the traditional quantum kernel. Moreover, better kernels will have flatter eigenspectra with more non-trivial kernel eigenvalues, which will lead to faster convergence speed \cite{roberts2021principles} (see discussions in SM) and less generalization error for good enough alignments \cite{bordelon2020spectrum,canatar2021spectral,simon2021neural,bahri2021explaining,atanasov2021neural}.

In Fig.~\ref{fig:kerstat}, we study the change of the kernel eigenspectra when turning on the Kerr coupling. We measure the complexity of the kernel by looking at the kernel effective dimension (the number of eigenvalues that are not small, where we set the criterion to be $>10^{-7}$). Since the kernel is normalized by the diagonal matrix element to be no larger than 1, we define the criterion to be $10^{-7}$ with respect to the norm of the kernel, which is also the numerical accuracy of our calculations. As shown in Fig 1, we find that when we turn on the Kerr coupling towards the non-perturbative regime, the kernel spectra have the tendency to be flatter, inferred from the increasing kernel effective dimension. This suggests that better performance can be achieved in generic numerical optimization experiments for larger Kerr non-linearity. In fact, a flatter spectrum means that the distribution has longer tails towards the higher values of the NTK eigenvalue parameters, which might indicate faster convergence and less generalization error for sufficient kernel alignments, the inner product between kernel predictions and the target functions \cite{bordelon2020spectrum,canatar2021spectral,simon2021neural,bahri2021explaining,atanasov2021neural}.

\begin{figure}[t]
\centering
\includegraphics[width=0.42\textwidth]{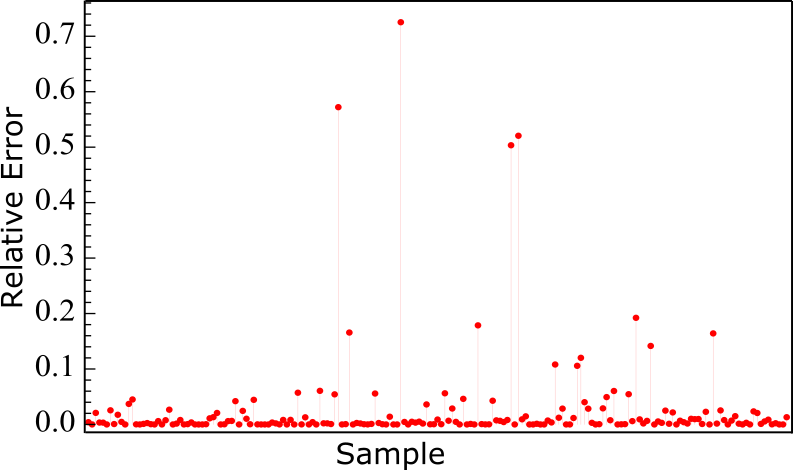}
\caption{The statistics of the relative error of matrix elements $\left| {({K_\mathrm{t}} - {K_\mathrm{e}})/{K_\mathrm{e}}} \right|$ for the perturbation theory prediction of the kernel ($K_\mathrm{t}$) and the numerical simulation ($K_\mathrm{e}$) among $\mathcal{O}(100)$ randomly generated data inputs $\mathbf{x}$ (the red points). We set $K_{\text{err}}=\mathcal{O}(0.01)\,\text{MHz}\times 2\pi$. }
\label{fig:pert}
\end{figure}

Analyses of kernel properties suggest that a non-zero $K_{\text{err}}$ might generically lead to better performance in supervised learning tasks. Our detailed perturbation analysis for leading order of $K_{\text{err}}$, shown in Supplementary Materials, is well-tested through numerics and provides reasonable predictions in the perturbative regime of $K_{\text{err}}$. The perturbation analysis corroborates our findings from numerical simulations -- analytic formulas obtained from the analysis match with numerical simulations of our bosonic quantum systems up to a given truncation. In Fig. \ref{fig:pert}, we verify the validity of our perturbation theory prediction by comparing it with the numerical simulation for truncated Hilbert space dimensions. $K_t$ and $K_e$ represent the value of matrix elements obtained from theory and simulations respectively. The relative error is negligible in most cases.

\emph{Supervised Learning and Gradient Descent.---}
We next focus on an actual machine learning task to discuss and demonstrate the quantum enhancement achieved by Kerr non-linearity. Consider a simple support vector machine model where $(\mathbf{x}_{\tilde{\alpha}},y_{\tilde{\alpha}})$ represent the corresponding input-output pairs. $\tilde{\alpha}$ denotes the index of data from the training set $\mathcal{A}$. In a supervised learning setting, while training, a predictive model is developed using $(\mathbf{x}_{\tilde{\alpha}},y_{\tilde{\alpha}})$. Subsequently, predictions $({z_\delta })$ can be obtained using this model:
\begin{align}
{z_\delta } = z (\mathbf{x}_\delta)=\sum\limits_{\tilde \alpha  \in \mathcal{A}} {{\theta _{\tilde \alpha }}K({\mathbf{x}_{\tilde \alpha }},{\mathbf{x}_\delta })} ~,
\end{align}
where $K$ is the kernel, $\mathbf{x}_\delta$ is a general input and $\delta$ denotes the index of data from the whole input data set $\mathcal{D}$ (so we have $\mathcal{A}\subset \mathcal{D}$), and $\theta_{\tilde{\alpha}}$ is a trainable variable. Ideally, in a quantum kernel method \cite{havlivcek2019supervised}, $K$ is evaluated using  quantum measurements, but here we will utilize numerical simulations and theoretical arguments to demonstrate quantum enhancement achieved by the Kerr kernel.

For our randomly generated $\mathbf{x}$, we assign them with one-dimensional outputs $y(\mathbf{x})$. The residual training error $\varepsilon (\mathbf{x}) = z (\mathbf{x}) -y (\mathbf{x}) $, where $ z (\mathbf{x}) $ is the kernel method prediction, enters in the formula of the mean-square loss function $\mathcal{L}=\frac{1}{2}\sum _{\mathbf{x}} \varepsilon^2 (\mathbf{x}) $. We use the gradient descent algorithm to minimize $\mathcal{L}$. In the kernel method, the optimization process is exactly solvable, and the decay of the residual training error is exponential when the learning rate is small, and the decay rate can be predicted by kernel eigenvalues (see SM). 

\begin{figure}[t]
\centering
\includegraphics[width=0.41\textwidth]{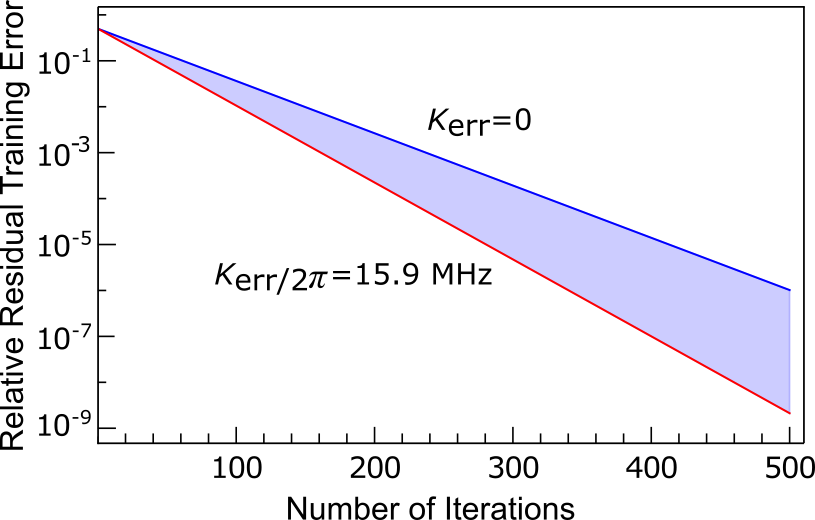}
\caption{We perform 30 different gradient descent processes with a fixed learning rate $10^{-3}$, by logarithmically increasing $K_{\text{err}}$ from 0 to $\mathcal{O}(10^3)\, \text{MHz}\times 2\pi$. The plots show the relative residual training error $\abs{\varepsilon (t)/\varepsilon (0)}$ depending on the iteration step $t$ (0 to 500), in the eigenvector direction of $K$ with the largest kernel eigenvalue. Theory (see SM) shows that this plot is independent of the choices of the supervised learning label $y$.}
\label{fig:op}
\end{figure}

In Fig.~\ref{fig:op}, we perform different gradient descent processes for increasing $K_{\text{err}}$. We find that the increasing Kerr non-linearity will generically accelerate the gradient descent dynamics, consistent with our findings of kernel eigenspectra (see SM) .

\begin{figure}[t]
\centering
\includegraphics[width=0.35\textwidth]{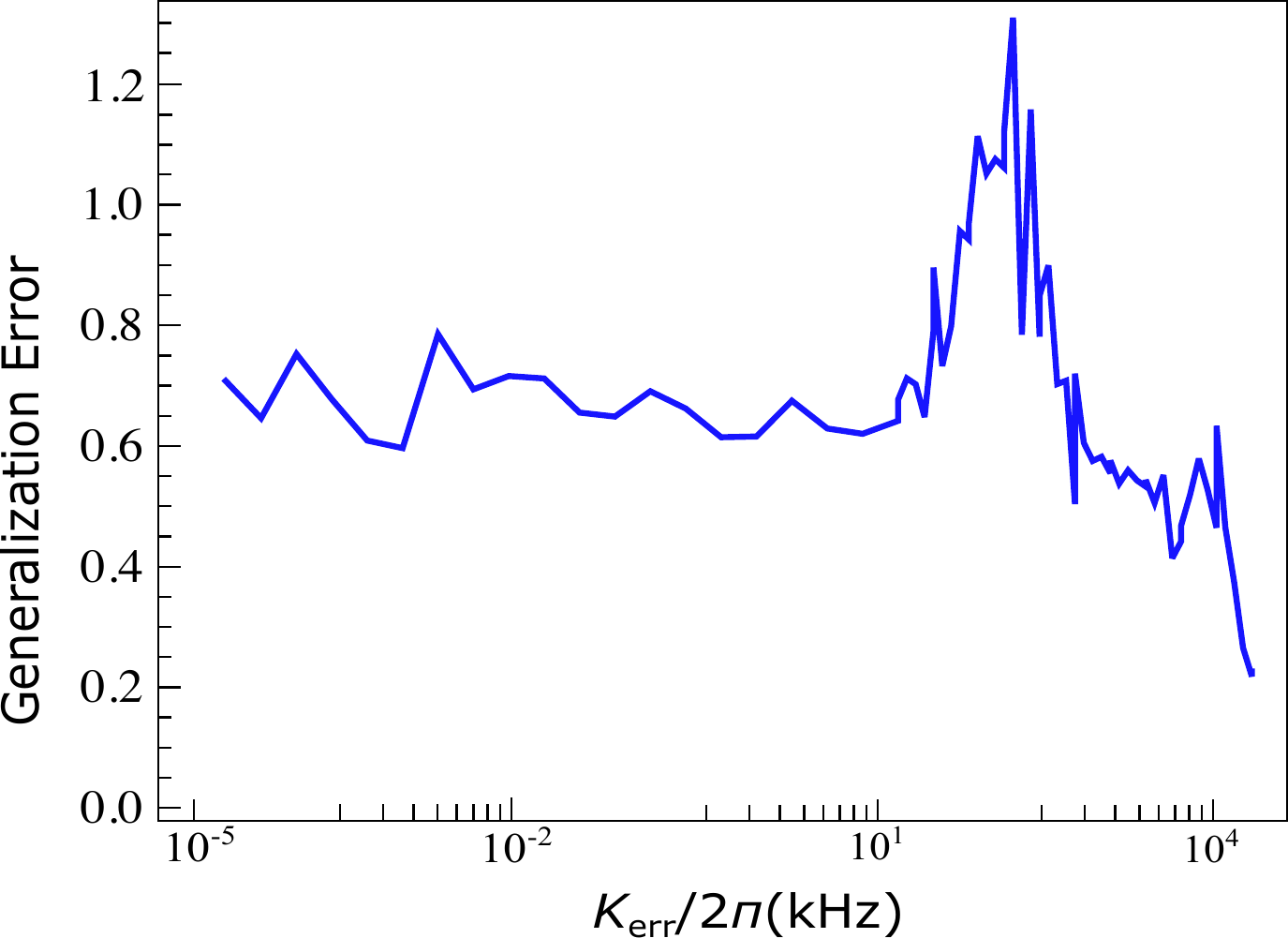}
\caption{Evaluating the algorithm performance through the generalization error. We set the training set
$\mathcal{T}$ to be the first half of the previous training set $\mathcal{A}$, and we denote the second half as the test set $\mathcal{B}$. We evaluate the generalization error using $\mathcal{L}_\mathcal{B}=\frac{1}{2\abs{\mathcal{B}}}\sum_{\mathbf{x} \in \mathcal{B}}\varepsilon(\mathbf{x})^2$ for different Kerr coefficients. Noises with standard deviation 0.1 is introduced in the label $y$ of the test set $\mathcal{B}$.} 
\label{fig:gene}
\end{figure}

\emph{Generalization Error.---}Another important metric for quantifying the performance of machine learning models is the generalization error. As an example, we split half of the data as the training set, and we evaluate the loss function on the remaining half as the test set after we train the model. We consider learning a function  $y (\mathbf{x})=\sum_{i=1}^3 \sin^2 (\mathbf{x}_i^2)$ instead, where our time $T$ is rescaled to be $\mathcal{O}(1)$ and other components in $T$ are set by dimensional analysis accordingly. Noise with the standard deviation 0.1 is introduced in all the instances of the task labels in $\mathcal{B}$. In Fig. \ref{fig:gene}, we find that generalization error has non-trivial behavior for growing Kerr non-linearity, and for large Kerr values in the non-perturbative regime, it decays significantly. According to Refs. \onlinecite{bordelon2020spectrum,canatar2021spectral,simon2021neural,bahri2021explaining,atanasov2021neural}, generalization errors could be related to neural tangent kernel eigenvalues, and the kernel method is generalized well from good enough alignments. Thus, our finding provides good evidence that quantum Kerr learning will provide extra enhancement in  algorithm performances. 

Moreover, it might be worth noticing that other studies \cite{banchi2021generalization,caro2022generalization} show related quantitative methods for predicting the generalization error for quantum machine learning methods. In \cite{banchi2021generalization}, it is shown that higher entropies in the kernel spectra might lead to worse generalizations in quantum machine learning. Our results are indeed consistent with such a theory, since the distribution of the kernel eigenvalues will lead to better generalization errors only when one has a good alignment between the target and the data. The kernel alignment \cite{cristianini2001kernel} is an inner product between the kernel in our assumption and the target function we want to fit and higher alignment will indicate better performance in generalization \cite{cristianini2001kernel,bordelon2020spectrum}. Intuitively, one could understand it as the following. If we have bad alignments between the target function we want to fit and the data, then the loss function landscape will be sharp for large kernel eigenvalues (especially when we consider the mean square error, and in this case, the (non-zero) neural tangent kernel eigenspectra are the same as the (non-zero) Hessian spectra, as shown by the singular value decomposition). Otherwise, flatter higher eigenvalues will lead to better generalizations, since there is more space to move around the local minima. Perhaps the condition among neural tangent kernels, alignments, and generalization errors could be refined by entropies, which is closely related to a potential information bottleneck theory in quantum machine learning \cite{toappear}. Finally, we emphasize that in our simulation, we do not consider quantum noises during sampling. However, one can still speak about the noise due to the randomness of the sampling from the input data distribution. More details are presented in the SM.

\emph{Higher dimensions.---} Regression in higher dimensions can be performed by using two quantum systems instead of one, each encoding a subset of the data,
\begin{align}
\begin{split}
    \mathcal{K} =\, & \bigg| \bigg\langle \psi \otimes \phi \bigg|
    \bar{\mathbb{T}}\exp \left[ {\frac{i}{\hbar }\textstyle{\int_0^{T'} {H(t')} dt'}} \right] \\
    & \times 
    \mathbb{T}\exp \left[  - \frac{i}{\hbar }\textstyle{\int_0^T {H(t)} dt} \right]\bigg|\psi \otimes \phi
    \bigg\rangle\bigg|^2
\end{split}~,
\end{align}
where we have assumed that we initially start in a product state. 
$H(t)$ could, again, be a general Hamiltonian. If $H(t)$ includes 
terms that couple the two subsystems, $|\psi\rangle$ and $|\phi\rangle$, this 
could potentially create entanglement between the two systems, increasing
the complexity and making it harder for a classical computer to compute. 
It is also interesting to look at the case when $H(t)$ does not couple
the two subsystems. For example, the Hamiltonian could represent two uncoupled 
Kerr resonators as
\begin{equation}
    H(t) = H\otimes I + I \otimes H,
\end{equation}
where $H$ is the Hamiltonian described in Equation~\eqref{hamiltonian} and $I$ 
is identity. In this initially
unentangled and uncoupled example, our kernel reduces to a Hadamard product 
of the two uncoupled kernels,
\begin{equation}
    \mathcal{K} = K_\alpha \circ K_\beta,
\end{equation}
where $K_\alpha$ ($K_\beta$) is the single-system Kerr kernel discussed in previous sections, and the subscript ($\alpha, \beta$) represents two independent subsystems. Because kernels are symmetric, positive definite matrices, we can bound the properties of
the eigenvalues of $\mathcal{K}$ using the spectra of $K$. For example, we have the following bound on the spectral radius,
$\rho$, of the product kernel~\cite{guo2019some} 
\begin{equation}
\rho(\mathcal{K}) \le \rho(K_\alpha)\rho(K_\beta).
\end{equation}
This bound is known to be rather loose, and tighter bounds
have been derived~\cite{guo2019some}.
As shown in Fig.~\ref{fig:kerstat}, the maximum kernel eigenvalues
of the single system grow with increasing $K_\mathrm{err}$. For 
multiple unentangled and uncoupled systems, we can expect that 
the maximal eigenvalues of this product kernel grow 
faster, leading to increased performance in 
higher-dimensional data sets compared with the non-Kerr kernel.
This can be extended to the product of many kernels, allowing
for the learning of arbitrary dimensional data with collections of single
oscillators.
It is likely that the addition of entanglement, through a Hamiltonian
term that couples the various systems, will further increase the 
performance~\cite{otten2020quantum}.

\emph{Conclusion and outlooks.---}Our paper opens up a novel direction by exploring the potential of quantum machine learning through circuit QED devices with non-trivial Kerr non-linearity. We find theoretical and numerical evidence where non-trivial Kerr coupling could significantly enhance the performance of the quantum kernel method, based on solid evaluations of kernel statistics and numerical optimization experiments. We strengthen our claims by applying the neural tangent kernel theory in machine learning, arguments from the theory of quantum complexity, and generalizations towards higher dimensions. Here, we suggest the following directions for future research.

a. Experimental implementation. It will be interesting to implement the proposal in our paper in the laboratory. When working with actual hardware, measurement noise and errors need to be considered. A combined design between theory and experiments could be obtained based on the theory of the kernel method, including trade-offs among the learning rate, the experimental precision for estimating the residual training error $\varepsilon$, and the number of gradient descent steps we could perform in the laboratory. 

b. Extensions to multiple modes. One could generalize the above work to multiple modes \cite{Han2016,Han2022}, where we expect the coupling between different bosonic modes will strengthen the complexity of the kernel. It will be interesting to see if our approach will realize the power of multimode devices towards hard problems in quantum machine learning.

c. Theoretical considerations. It will be interesting to explore further the complexity foundations of our claims. The argument about the Kerr non-linearity and complexity, although promising, is not proven rigorously. A more solid statement about complexity might deepen our understanding of the algorithmic potential of the Kerr non-linearity.

\begin{acknowledgments}
We thank David Meltzer, Daniel A. Roberts and Quntao Zhuang for useful discussions. We thank Liang Jiang for his numerous helpful suggestions. Work performed at the Center for Nanoscale Materials, a U.S. Department of Energy Office of Science User Facility, was supported by the U.S. DOE, Office of Basic Energy Sciences, under Contract No. DE-AC02-06CH11357.
JL is supported in part by International Business Machines (IBM) Quantum through the Chicago Quantum Exchange, and the Pritzker School of Molecular Engineering at the University of Chicago through AFOSR MURI (FA9550-21-1-0209). JL also serves as a scientific advisor of qBraid Co.. CZ and JY acknowledges support from the ARO (W911NF-18-1-0020, W911NF-18-1-0212), ARO MURI (W911NF-16-1-0349), AFOSR MURI (FA9550-19-1-0399, FA9550-21-1-0209), DoE Q-NEXT, NSF (EFMA-1640959, OMA-1936118, EEC-1941583), NTT Research, and the Packard Foundation (2013-39273).
\end{acknowledgments}

\bibliographystyle{apsrev4-1}
\bibliography{singlemode.bib}

\pagebreak
\clearpage
\foreach \x in {1,...,\the\pdflastximagepages}
{
	\clearpage
	

\section*{Supplementary material}

# Supplementary Material: Quantum Kerr Learning

## CONTENTS

## I. KERNEL METHOD AND THE GRADIENT DESCENT DYNAMICS

Here we provide a self-contained introduction to the basic methodology we use. For more details about kernel methods and neural tangent kernels, see [1–3].

We consider a simple supportive vector machine model. We start from a series of data $(x_{\tilde{\alpha}}, y_{\tilde{\alpha}})$, where $\tilde{\alpha}$ denotes the index of data from the training set $\mathcal{A}$, and we are in a supervised learning task with the output label $y_{\tilde{\alpha}}$. We wish to fit the answer with the linear model,

$$z_\delta = z(x_\delta) = \sum_{\tilde{\alpha} \in \mathcal{A}} \theta_{\tilde{\alpha}} K(x_{\tilde{\alpha}}, x_\delta) , \tag{S1}$$

where $K$ is the kernel, $x_\delta$ is a general input and $\delta$ denotes the index of data from the whole input data set $\mathcal{D}$ (so we have $\mathcal{A} \subset \mathcal{D}$), and $\theta_\delta$ is a trainable variable. In the quantum kernel method [4], $K$ is evaluated using the quantum measurement. We set the loss function

$$L_{\mathcal{A}} = \frac{1}{2} \sum_{\tilde{\alpha} \in \mathcal{A}} (z_{\tilde{\alpha}} - y_{\tilde{\alpha}})^2 = \frac{1}{2} \sum_{\tilde{\alpha} \in \mathcal{A}} \varepsilon_{\tilde{\alpha}}^2 . \tag{S2}$$

Here, $\varepsilon$ is the residual training error. We consider the simplest gradient descent dynamics,

$$\delta\theta_{\tilde{\alpha}} = \theta_{\tilde{\alpha}}(t+1) - \theta_{\tilde{\alpha}}(t) = -\eta \frac{dL_{\mathcal{A}}}{d\theta_{\tilde{\alpha}}} = -\eta \sum_{\tilde{\beta} \in \mathcal{A}} \frac{d\varepsilon_{\tilde{\beta}}}{d\theta_{\tilde{\alpha}}} \varepsilon_{\tilde{\beta}} . \tag{S3}$$

Moreover, since our model is linear, we have

$$\delta\varepsilon_{\tilde{\alpha}} = \sum_{\tilde{\beta} \in \mathcal{A}} \frac{d\varepsilon_{\tilde{\alpha}}}{d\theta_{\tilde{\beta}}} \delta\theta_{\tilde{\beta}} = -\eta \sum_{\tilde{\beta}, \tilde{\gamma} \in \mathcal{A}} \frac{d\varepsilon_{\tilde{\alpha}}}{d\theta_{\tilde{\beta}}} \frac{d\varepsilon_{\tilde{\gamma}}}{d\theta_{\tilde{\beta}}} \varepsilon_{\tilde{\gamma}} . \tag{S4}$$

We define

$$K_{H,\tilde{\alpha}\tilde{\gamma}} = \sum_{\tilde{\beta} \in \mathcal{A}} \frac{d\varepsilon_{\tilde{\alpha}}}{d\theta_{\tilde{\beta}}} \frac{d\varepsilon_{\tilde{\gamma}}}{d\theta_{\tilde{\beta}}} . \tag{S5}$$

So we get

$$\delta\varepsilon_{\tilde{\alpha}} = \sum_{\tilde{\beta} \in \mathcal{A}} \frac{d\varepsilon_{\tilde{\alpha}}}{d\theta_{\tilde{\beta}}} \delta\theta_{\tilde{\beta}} = -\eta \sum_{\tilde{\gamma} \in \mathcal{A}} K_{H,\tilde{\alpha}\tilde{\gamma}} \varepsilon_{\tilde{\gamma}} . \tag{S6}$$

The quantity $K_H$ is called the neural tangent kernel (NTK) [3, 5–11]. In the supportive vector machine case, $K_H$ is a constant,

$$\frac{d\varepsilon_{\tilde{\alpha}}}{d\theta_{\tilde{\beta}}} = K(x_{\tilde{\beta}}, x_{\tilde{\alpha}}) \,,$$

$$K_{H,\tilde{\alpha}\tilde{\gamma}} = \sum_{\tilde{\beta}\in\mathcal{A}} \frac{d\varepsilon_{\tilde{\alpha}}}{d\theta_{\tilde{\beta}}} \frac{d\varepsilon_{\tilde{\gamma}}}{d\theta_{\tilde{\beta}}} = \sum_{\tilde{\beta}\in\mathcal{A}} K(x_{\tilde{\beta}}, x_{\tilde{\alpha}}) K(x_{\tilde{\beta}}, x_{\tilde{\gamma}}) = \sum_{\tilde{\beta}\in\mathcal{A}} K(x_{\tilde{\alpha}}, x_{\tilde{\beta}}) K(x_{\tilde{\beta}}, x_{\tilde{\gamma}}) \,. \tag{S7}$$

Here, we have used the fact that $K$ is symmetric. If we use the compact matrix notation in the sample space, we have

$$K_H = K^2 \,. \tag{S8}$$

The solution is analytically solved since $K_H$ or $K$ are constants,

$$\varepsilon_{\tilde{\alpha}}(t) = \left( \left(1 - \eta K_H\right)^t \right)_{\tilde{\alpha}\tilde{\beta}} \varepsilon_{\tilde{\beta}}(0) \,. \tag{S9}$$

The properties of matrix $H = K^2$ would affect the efficiency and performance of the gradient descent. Say that we diagonalize the matrix $K$ by

$$K_{\tilde{\alpha}\tilde{\beta}} = (V\mathbb{K}V^T)_{\tilde{\alpha}\tilde{\beta}} \,. \tag{S10}$$

Here $V$ is the matrix of eigenvectors of $K$, and $\mathbb{K}$ is the matrix $K$ in the eigenspace,

$$\mathbb{K} = \mathrm{diag}\left(\kappa_{\tilde{\alpha}}\right) \,,$$
$$\mathbb{K}_{\tilde{\alpha}\tilde{\beta}} = \kappa_{\tilde{\alpha}}\delta_{\tilde{\alpha}\tilde{\beta}} \,. \tag{S11}$$

We have

$$\varepsilon_{\tilde{\alpha}}(t) = \left( V(1 - \eta\mathbb{K}^2)^t V^T \right)_{\tilde{\alpha}\tilde{\beta}} \varepsilon_{\tilde{\beta}}(0) \,. \tag{S12}$$

We define

$$u_{\tilde{\alpha}} = V^T_{\tilde{\alpha}\tilde{\beta}} \varepsilon_{\tilde{\beta}} \,, \tag{S13}$$

so we get

$$u_{\tilde{\alpha}}(t) = (1 - \eta\kappa_{\tilde{\alpha}}^2)^t u_{\tilde{\alpha}}(0) \,. \tag{S14}$$

Thus the spectrum of $K$ gives the characteristic decaying time of the residual training error in the inertial frame, $u$. More precisely, in the small $\eta$ limit, we get

$$u_{\tilde{\alpha}}(t) \approx e^{-\eta\kappa_{\tilde{\alpha}}^2 t} u_{\tilde{\alpha}}(0) \,. \tag{S15}$$

Thus, a larger absolute value of $\kappa_{\tilde{\alpha}}$ will lead to a faster decay rate. If our kernel is Gaussian, there will be a sufficiently large amount of eigenvalues accumulated around zero, because of the long tails of the distribution. Thus, if our spectra of $K_H$ or $K$ are flatter, our performance is expected to be better.

Finally, we make a short discussion about the generalization error. For a general input data $\mathbf{x}_\delta$, the output at the end of training is given by [2, 3],

$$z_\delta(\infty) = z_\delta(0) - \sum_{\tilde{\alpha},\tilde{\alpha}_1\in\mathcal{A}} H_{\delta\tilde{\alpha}}\tilde{H}^{\tilde{\alpha}\tilde{\alpha}_1}\varepsilon_{\tilde{\alpha}_1}(0) \,. \tag{S16}$$

Here, we use the notation where $\tilde{H}^{\tilde{\alpha}_1\tilde{\alpha}_2} = (H)^{-1}_{\tilde{\alpha}_1\tilde{\alpha}_2}$, the inverse of the matrix $H$ restricted at the subspace $\mathcal{A} \subset \mathcal{D}$. For example, in the main text in practice we take the initial trainable weights $W(0) = 0$, so we get

$$z_\delta(\infty) = \sum_{\tilde{\alpha},\tilde{\alpha}_1\in\mathcal{A}} H_{\delta\tilde{\alpha}}\tilde{H}^{\tilde{\alpha}\tilde{\alpha}_1} y_{\tilde{\alpha}_1} \,. \tag{S17}$$

Specifically, if we take a data point inside the space $\mathcal{A}$, we get

$$z_{\tilde{\delta}}(\infty) = \sum_{\tilde{\alpha},\tilde{\alpha}_1\in\mathcal{A}} H_{\tilde{\delta}\tilde{\alpha}}\tilde{H}^{\tilde{\alpha}\tilde{\alpha}_1} y_{\tilde{\alpha}_1} = \sum_{\tilde{\alpha}_1\in\mathcal{A}} \delta_{\tilde{\delta}}^{\tilde{\alpha}_1} y_{\tilde{\alpha}_1} = y_{\tilde{\delta}} \,, \tag{S18}$$

by the definition of matrix inverse. One could define the generalization error as

$$\mathcal{L}_\mathcal{B} = \frac{1}{2|\mathcal{B}|} \sum_{\beta\in\mathcal{B}} (z_\beta - y_\beta)^2 = \frac{1}{2|\mathcal{B}|} \sum_{\beta\in\mathcal{B}} \varepsilon_\beta^2 \,. \tag{S19}$$

Here, $\mathcal{B} \subset \mathcal{D}$ is a test set and $\mathcal{B} \neq \mathcal{A}$.

## II.    COMPLEXITY THEORY STATEMENTS

Here, we give some complexity-theory-related arguments about the machine learning potential of non-linear Kerr terms. In fact, a general real-time evolution of a quantum system could be formulated in the complexity class **BQP**, which is the class of problems solved by quantum computers with bounded error and polynomial time. More precisely, the problem is **BQP**-complete. As long as **BQP** is not equal to **P**, the problem will be hard to simulate in general for classical computers [12–14]. In the bosonic system, the quadratic Hamiltonian is simulatable by classical computers in polynomial time, based on the celebrated Gottesman-Knill theorem [15] in the bosonic case with the Gaussian initial states [16]. Thus, since our Hamiltonian includes Kerr non-linearity, by the inverse statement of the bosonic Gottesman-Knill theorem, we suggest that our Hamiltonian Eq. (**??**) might be challenging to simulate for classical computers for large scale simulations, and for our analog quantum system, one could evaluate the kernel efficiently within the polynomial time.

Moreover, we compare the above statements with the paper [17] addressing complexity-related statements. in [17], the paper gives examples where kernels could be computationally hard. There are illustrations on the complexity theory (see Figure 1 of [17]), which shows computational complexity where theories with data could be located. However, their proofs are based on the statistical learning theory, which could prove statements in the PAC-learning framework. Those theories (see [17], and also [18]) only give loose lower bounds on the learnability without addressing any actual algorithms about kernel methods or gradient descent. Algorithm performances are not considered in such statistical learning theory frameworks. Thus, we believe that kernels existing in the computational class **BQP** should give quantum enhancements and possible advantages towards classical counterparts, just like examples illustrated in [19], which are also kernel methods, hard to simulate classically, satisfying but significantly beyond the statistical learning theory lower bounds.

## III.    THE HAMILTONIAN AND THE PERTURBATION THEORY

### A.    Problem setup

We deal with the Hamiltonian,

$$H = H_0 + H_t + H_I = H_f + H_I \,, \tag{S20}$$

where

$$\begin{aligned}
\frac{H_0}{\hbar} &= \omega_m \left( b^\dagger b + \frac{1}{2} \right) \,, \\
\frac{H_t}{\hbar} &= \Omega \left( b e^{-i\omega_L t} + b^\dagger e^{i\omega_L t} \right) \,, \\
\frac{H_I}{\hbar} &= K_{\mathrm{err}} b^\dagger b^\dagger b b \,.
\end{aligned} \tag{S21}$$

Now we wish to evaluate

$$K = \left| \left\langle \psi \left| \left( \overline{\mathbb{T}} \exp\left[ \frac{i}{\hbar} \int_0^T H(\Omega, \omega_L) dt \right] \times \mathbb{T} \exp\left[ -\frac{i}{\hbar} \int_0^{T'} H(\Omega', \omega_L') dt \right] \right) \right| \psi \right\rangle \right|^2 \,. \tag{S22}$$

Note that we wish $\omega_m$ to be fixed.

### B.    Free theory

Now we use the rotation frame method to simplify the problem. For an arbitrary quantum system generated by

$$i\hbar \frac{d}{dt} \left| \phi(t) \right\rangle = H(t) \left| \phi(t) \right\rangle \,. \tag{S23}$$

We assume a rotation frame such that

$$\left| \psi(t) \right\rangle = U_R(t) \left| \phi(t) \right\rangle \,. \tag{S24}$$

We get

$$i\hbar \frac{d}{dt} \left| \psi(t) \right\rangle = H_R(t) \left| \psi(t) \right\rangle \,, \tag{S25}$$

where

$$H_R(t) = i\hbar \dot{U}_R(t)U^\dagger(t) + U_R(t)H(t)U_R^\dagger(t) \ . \tag{S26}$$

$H_R$ is, in a sense, a rotated definition of the Hamiltonian. Without Kerr non-linearity, we could apply it to

$$H_f(t) = H_0 + H_t(t) = \hbar\omega_m \left( b^\dagger b + \frac{1}{2} \right) + \hbar\Omega \left( be^{-i\omega_L t} + b^\dagger e^{i\omega_L t} \right) \ , \tag{S27}$$

with

$$U_R(t) = e^{i\omega_L t b^\dagger b} \ . \tag{S28}$$

So we have

$$\begin{aligned}
H_R(t) &= i\hbar \dot{U}_R(t)U_R^\dagger(t) + U_R(t)H_f(t)U_R^\dagger(t) \\
&= \hbar\left(\omega_m - \omega_L\right)\left(b^\dagger b\right) + \frac{1}{2}\hbar\omega_m + \hbar\Omega(b + b^\dagger) \ .
\end{aligned} \tag{S29}$$

In fact, we find $H_R$ is static. Now, remember that our feature map is

$$|\phi(\mathbf{x})\rangle = \mathbb{T}\exp\left[ -\frac{i}{\hbar}\int_0^T H_f dt \right] |\psi\rangle \ , \tag{S30}$$

which is driven by the Hamiltonian,

$$\frac{d |\phi(\mathbf{x})\rangle}{dT} = -\frac{i}{\hbar}H_f(T) |\phi(\mathbf{x})\rangle \ . \tag{S31}$$

So we define

$$|\psi(\mathbf{x})\rangle = U_R(T) |\phi(\mathbf{x})\rangle \ , \tag{S32}$$

and we have

$$|\psi(\mathbf{x})\rangle = \mathbb{T}\exp\left[ -\frac{i}{\hbar}\int_0^T H_R dt \right] |\psi\rangle = \exp\left[ -\frac{i}{\hbar}H_R T \right] |\psi\rangle \ . \tag{S33}$$

Basically, we just prove,

$$U_R(T)\mathbb{T}\exp\left[ -\frac{i}{\hbar}\int_0^T H_f dt \right] = \exp\left[ -\frac{i}{\hbar}H_R T \right] \ . \tag{S34}$$

Now, we could write down the kernel formula as

$$K(\mathbf{x}, \mathbf{x}') = |\langle\phi(\mathbf{x})|\phi(\mathbf{x}')\rangle|^2 = \left| \left\langle \psi(\mathbf{x})|U_R(\mathbf{x})U_R^\dagger(\mathbf{x}')|\psi(\mathbf{x}') \right\rangle \right|^2 \ . \tag{S35}$$

Here we denote

$$U_R(\mathbf{x}) = U_R(T) = \exp\left( i\omega_L T b^\dagger b \right) \ . \tag{S36}$$

Now we start to compute $\psi(\mathbf{x})$. We have

$$\begin{aligned}
H_R &= \hbar\left(\omega_m - \omega_L\right)\left(b^\dagger b\right) + \frac{1}{2}\hbar\omega_m + \hbar\Omega\left(b + b^\dagger\right) \\
&= -\hbar\Delta_L\left(b^\dagger b\right) + \hbar\Omega\left(b + b^\dagger\right) + \frac{1}{2}\hbar\omega_m \\
&= \hbar Q_L\left(b^\dagger b\right) + \hbar\Omega\left(b + b^\dagger\right) + \frac{1}{2}\hbar\omega_m \ ,
\end{aligned} \tag{S37}$$

where we define $\Delta_L = \omega_L - \omega_m = -Q_L$. We could perform the transformation,

$$b = a - \lambda \, ,$$
$$b^\dagger = a^\dagger - \lambda \, ,$$

(S38)

where we assume that $\lambda$ is real. In this case we have

$$
\begin{aligned}
b^\dagger b &= a^\dagger a - \lambda (a^\dagger + a) + \lambda^2 \\
&= a^\dagger a - \lambda (b^\dagger + b + 2\lambda) + \lambda^2 \\
&= a^\dagger a - \lambda (b^\dagger + b) - \lambda^2 \, .
\end{aligned}
$$

(S39)

We notice that we have $[a, a^\dagger] = 1$. And we have

$$
\begin{aligned}
H_R &= \hbar Q_L \left( b^\dagger b \right) + \hbar \Omega \left( b + b^\dagger \right) + \frac{1}{2} \hbar \omega_m \\
&= \hbar Q_L a^\dagger a - \lambda \hbar Q_L \left( b + b^\dagger \right) - \hbar Q_L \lambda^2 + \hbar \Omega \left( b + b^\dagger \right) + \frac{1}{2} \hbar \omega_m \, .
\end{aligned}
$$

(S40)

We set

$$\lambda Q_L = \Omega \, .$$

(S41)

So we have

$$H_R = \hbar Q_L a^\dagger a - \hbar Q_L \lambda^2 + \frac{1}{2} \hbar \omega_m \, .$$

(S42)

For simplicity, we assume that the initial state is expanded in the coherent state basis,

$$|\psi\rangle = \sum_\alpha f(\alpha) |\alpha\rangle_b \, ,$$

(S43)

where $|\alpha\rangle_b$ is the coherent state corresponding to the creation operator $b^\dagger$. Note that

$$|\alpha\rangle_b \equiv |\alpha'\rangle_a = |\alpha + \lambda\rangle_a \, ,$$

(S44)

since

$$
\begin{aligned}
b|\alpha\rangle_b &= (a - \lambda)|\alpha\rangle_b = (\alpha' - \lambda)|\alpha\rangle_b = \alpha|\alpha\rangle_b \, , \\
a|\alpha\rangle_b &= a|\alpha'\rangle_a = \alpha'|\alpha'\rangle_a = (\alpha + \lambda)|\alpha\rangle_b \, .
\end{aligned}
$$

(S45)

The above equation is not precise, since the equation should be up to a global factor. A more precise calculation shows,

$$
\begin{aligned}
|\alpha + \lambda\rangle_a &= e^{(\alpha+\lambda)a^\dagger - (\alpha^*+\lambda)a}|0\rangle_a = e^{(\alpha+\lambda)a^\dagger - (\alpha^*+\lambda)a} e^{-\lambda b^\dagger + \lambda b}|0\rangle_b \\
&= e^{(\alpha+\lambda)(b^\dagger+\lambda) - (\alpha^*+\lambda)(b+\lambda)} e^{-\lambda b^\dagger + \lambda b}|0\rangle_b \, ,
\end{aligned}
$$

(S46)

since,

$$
\begin{aligned}
&\left[ (\alpha+\lambda)(b^\dagger + \lambda) - (\alpha^*+\lambda)(b+\lambda), -\lambda b^\dagger + \lambda b \right] \\
&= -\lambda(\alpha - \alpha^*) \, .
\end{aligned}
$$

(S47)

Using the BCH formula, we get

$$|\alpha + \lambda\rangle_a = e^{(i\mathrm{Im}\alpha)\lambda}|\alpha\rangle_b \, .$$

(S48)

Thus,

$$
\begin{aligned}
|\alpha + \lambda\rangle_a &= e^{(i\mathrm{Im}\alpha)\lambda}|\alpha\rangle_b \, , \\
|\alpha\rangle_a &= e^{(i\mathrm{Im}\alpha)\lambda}|\alpha - \lambda\rangle_b \, .
\end{aligned}
$$

(S49)

So we have

$$|\psi\rangle = \sum_\alpha f(\alpha)|\alpha\rangle_b = \sum_\alpha f(\alpha)e^{-(i\mathrm{Im}\alpha)\lambda}|\alpha+\lambda\rangle_a = \sum_\alpha f(\alpha-\lambda)e^{-(i\mathrm{Im}\alpha)\lambda}|\alpha\rangle_a \ . \tag{S50}$$

Thus, we get

$$\exp\left[-\frac{i}{\hbar}H_R T\right]|\psi\rangle = \exp\left[-\frac{iT}{\hbar}\left(\hbar Q_L a^\dagger a - \hbar Q_L \lambda^2 + \frac{1}{2}\hbar\omega_m\right)\right]|\psi\rangle$$

$$= e^{-\frac{1}{2}i\omega_m T + iQ_L\lambda^2 T}\sum_{\alpha'} f(\alpha-\lambda)e^{-(i\mathrm{Im}\alpha)\lambda}\exp\left[-iQ_L T\left(a^\dagger a\right)\right]|\alpha\rangle_a$$

$$= e^{-\frac{1}{2}i\omega_m T + iQ_L\lambda^2 T}\sum_\alpha f(\alpha-\lambda)e^{-(i\mathrm{Im}\alpha)\lambda}\left|\alpha e^{-iQ_L T}\right\rangle_a$$

$$= e^{-\frac{1}{2}i\omega_m T + iQ_L\lambda^2 T}\sum_\alpha f(\alpha-\lambda)e^{i\mathrm{Im}(\alpha(e^{-iQ_L T}-1))\lambda}\left|\alpha e^{-iQ_L T}-\lambda\right\rangle_b$$

$$= e^{-\frac{1}{2}i\omega_m T + iQ_L\lambda^2 T}\sum_\beta f((\beta+\lambda)e^{iQ_L T}-\lambda)e^{i\lambda\mathrm{Im}\left[(\beta+\lambda)(1-e^{iQ_L T})\right]}|\beta\rangle_b$$

$$= e^{-\frac{1}{2}i\omega_m T + iQ_L\lambda^2 T + i\lambda\mathrm{Im}\left[\lambda(1-e^{iQ_L T})\right]}\sum_\beta f((\beta+\lambda)e^{iQ_L T}-\lambda)e^{i\lambda\mathrm{Im}\left[\beta(1-e^{iQ_L T})\right]}|\beta\rangle_b \ . \tag{S51}$$

So we have

$$U_R^\dagger|\psi(\mathbf{x})\rangle$$

$$= e^{-\frac{1}{2}i\omega_m T + iQ_L\lambda^2 T + i\lambda\mathrm{Im}\left[\lambda\left(1-e^{iQ_L T}\right)\right]}\sum_\beta f\left((\beta+\lambda)e^{iQ_L T}-\lambda\right)e^{i\lambda\mathrm{Im}\left[\beta\left(1-e^{iQ_L T}\right)\right]}e^{-i\omega_L T b^\dagger b}|\beta\rangle_b$$

$$= e^{-\frac{1}{2}i\omega_m T + iQ_L\lambda^2 T + i\lambda\mathrm{Im}\left[\lambda\left(1-e^{iQ_L T}\right)\right]}\sum_\beta f\left((\beta+\lambda)e^{iQ_L T}-\lambda\right)e^{i\lambda\mathrm{Im}\left[\beta\left(1-e^{iQ_L T}\right)\right]}\left|e^{-i\omega_L T}\beta\right\rangle_b$$

$$= e^{-\frac{1}{2}i\omega_m T + iQ_L\lambda^2 T - i\lambda^2\sin(Q_L T)}\sum_\beta f\left(\left(e^{i\omega_L T}\beta+\lambda\right)e^{iQ_L T}-\lambda\right)e^{i\lambda\mathrm{Im}\left[\beta e^{i\omega_L T}\left(1-e^{iQ_L T}\right)\right]}|\beta\rangle_b \ . \tag{S52}$$

We could re-write it by

$$\left(e^{i\omega_L T}\beta+\lambda\right)e^{iQ_L T}-\lambda = e^{i\omega_m T}\beta - \left(1-e^{iQ_L T}\right)\frac{\Omega}{Q_L} \ ,$$

$$\omega_L = \omega_m - Q_L \ . \tag{S53}$$

Now we visit a specific situation, where

$$|\psi\rangle = \sum_\alpha f(\alpha)|\alpha\rangle_b \Rightarrow \int d\alpha f(\alpha)|\alpha\rangle_b = \int d\alpha\delta(\alpha-\alpha_i)|\alpha\rangle_b = |\alpha_i\rangle_b \ , \tag{S54}$$

and we visit the case where $\alpha_i$ is real. So we impose

$$e^{i\omega_m T}\beta - \left(1-e^{iQ_L T}\right)\frac{\Omega}{Q_L} = \alpha_i \ . \tag{S55}$$

And we have

$$\beta = e^{-i\omega_m T}\alpha_i + \left(1-e^{iQ_L T}\right)\frac{\Omega}{Q_L}e^{-i\omega_m T} \ . \tag{S56}$$

So we get

$$U_R^\dagger|\psi(\mathbf{x})\rangle = e^{-\frac{1}{2}i\omega_m T + i\frac{\Omega^2}{Q_L}T - i\frac{\Omega(\alpha_i Q_L + \Omega)}{Q_L^2}\sin(Q_L T)}\left|e^{-i\omega_m T}\alpha_i + \left(1-e^{iQ_L T}\right)\frac{\Omega}{Q_L}e^{-i\omega_m T}\right\rangle_b$$

$$\propto \left|e^{-i\omega_m T}\alpha_i + \left(1-e^{iQ_L T}\right)\frac{\Omega}{Q_L}e^{-i\omega_m T}\right\rangle_b \ . \tag{S57}$$

### C.   Perturbation theory

We could use a similar trick to study the perturbation theory of the Kerr non-linearity. First, we notice that

$$
\begin{aligned}
H_I &= \hbar K_{\mathrm{err}} b^\dagger b^\dagger bb = \hbar K_{\mathrm{err}} b^\dagger bb^\dagger b + \hbar K_{\mathrm{err}} b^\dagger \left[ b^\dagger, b \right] b \\
&= \hbar K_{\mathrm{err}} b^\dagger bb^\dagger b - \hbar K_{\mathrm{err}} b^\dagger b = \hbar K_{\mathrm{err}} N^2 - \hbar K_{\mathrm{err}} N \ ,
\end{aligned}
\tag{S58}
$$

So we have

$$
H_R(t) = H_R = \hbar Q_L \left( b^\dagger b \right) + \frac{1}{2} \hbar \omega_m + \hbar K_{\mathrm{err}} b^\dagger b^\dagger bb + \hbar \Omega \left( b + b^\dagger \right) \ .
\tag{S59}
$$

Moreover, using the Feynman's disentangling formula,

$$
\exp \left( T(A + B) \right) = \exp \left( TA \right) \mathbb{T} \exp \left[ \int_0^T \tilde{B} \left( \tau \right) d\tau \right] \ ,
\tag{S60}
$$

where

$$
\tilde{B}(\tau) \equiv \exp \left( -\tau A \right) B \exp \left( \tau A \right) \ .
\tag{S61}
$$

So we get

$$
\begin{aligned}
&\exp \left[ -\frac{i}{\hbar} H_R T \right] |\psi\rangle \\
&= \exp \left[ -\frac{i}{\hbar} \left( \hbar Q_L \left( b^\dagger b \right) + \frac{1}{2} \hbar \omega_m + \hbar \Omega \left( b + b^\dagger \right) \right) T \right] \mathbb{T} \exp \left[ -\frac{i}{\hbar} \int_0^T \tilde{H}_I \left( \tau \right) d\tau \right] |\psi\rangle \\
&= e^{-\frac{i}{\hbar} \left( \hbar Q_L \left( b^\dagger b \right) + \frac{1}{2} \hbar \omega_m + \hbar \Omega \left( b + b^\dagger \right) \right) T} \left( 1 - \frac{i}{\hbar} \int_0^T \tilde{H}_I \left( \tau \right) d\tau + \mathcal{O}(K_{\mathrm{err}}^2) \right) |\psi\rangle \ ,
\end{aligned}
\tag{S62}
$$

where

$$
\tilde{H}_I(\tau) = \hbar K_{\mathrm{err}} e^{\frac{i}{\hbar} \left( \hbar Q_L \left( b^\dagger b \right) + \frac{1}{2} \hbar \omega_m + \hbar \Omega \left( b + b^\dagger \right) \right) \tau} b^\dagger b^\dagger bb \, e^{-\frac{i}{\hbar} \left( \hbar Q_L \left( b^\dagger b \right) + \frac{1}{2} \hbar \omega_m + \hbar \Omega \left( b + b^\dagger \right) \right) \tau} \ .
\tag{S63}
$$

We could write

$$
|\psi(\mathbf{x})\rangle = \exp \left[ -\frac{i}{\hbar} H_R T \right] |\psi\rangle = |\psi_0(\mathbf{x})\rangle + |\delta\psi(\mathbf{x})\rangle \ ,
\tag{S64}
$$

where

$$
|\psi_0(\mathbf{x})\rangle = e^{-\frac{i}{\hbar} \left( \hbar Q_L \left( b^\dagger b \right) + \frac{1}{2} \hbar \omega_m + \hbar \Omega \left( b + b^\dagger \right) \right) T} |\psi\rangle \ ,
\tag{S65}
$$

which has been computed before, and

$$
\begin{aligned}
|\delta\psi(\mathbf{x})\rangle &= -\frac{i}{\hbar} e^{-\frac{i}{\hbar} \left( \hbar Q_L \left( b^\dagger b \right) + \frac{1}{2} \hbar \omega_m + \hbar \Omega \left( b + b^\dagger \right) \right) T} \int_0^T \tilde{H}_I \left( \tau \right) |\psi\rangle \, d\tau = -i K_{\mathrm{err}} e^{-\frac{i}{\hbar} \left( \hbar Q_L \left( b^\dagger b \right) + \frac{1}{2} \hbar \omega_m + \hbar \Omega \left( b + b^\dagger \right) \right) T} \\
&\int_0^T e^{\frac{i}{\hbar} \left( \hbar Q_L \left( b^\dagger b \right) + \frac{1}{2} \hbar \omega_m + \hbar \Omega \left( b + b^\dagger \right) \right) \tau} b^\dagger b^\dagger bb \, e^{-\frac{i}{\hbar} \left( \hbar Q_L \left( b^\dagger b \right) + \frac{1}{2} \hbar \omega_m + \hbar \Omega \left( b + b^\dagger \right) \right) \tau} |\psi\rangle \, d\tau \ .
\end{aligned}
\tag{S66}
$$

Moreover, we have

$$
\begin{aligned}
\tilde{H}_I(\tau) &= \hbar K_{\mathrm{err}} e^{\frac{i}{\hbar} \left( \hbar Q_L \left( b^\dagger b \right) + \frac{1}{2} \hbar \omega_m + \hbar \Omega \left( b + b^\dagger \right) \right) \tau} b^\dagger b^\dagger bb \, e^{-\frac{i}{\hbar} \left( \hbar Q_L \left( b^\dagger b \right) + \frac{1}{2} \hbar \omega_m + \hbar \Omega \left( b + b^\dagger \right) \right) \tau} \\
&= \hbar K_{\mathrm{err}} e^{\frac{i}{\hbar} \left( \hbar Q_L \left( b^\dagger b \right) + \hbar \Omega \left( b + b^\dagger \right) \right) \tau} b^\dagger b^\dagger bb \, e^{-\frac{i}{\hbar} \left( \hbar Q_L \left( b^\dagger b \right) + \hbar \Omega \left( b + b^\dagger \right) \right) \tau} \\
&= \hbar K_{\mathrm{err}} e^{\frac{i}{\hbar} \left( \hbar Q_L a^\dagger a - \hbar Q_L \lambda^2 \right) \tau} b^\dagger b^\dagger bb \, e^{-\frac{i}{\hbar} \left( \hbar Q_L a^\dagger a - \hbar Q_L \lambda^2 \right) \tau} \\
&= \hbar K_{\mathrm{err}} e^{\frac{i}{\hbar} \left( \hbar Q_L a^\dagger a \right) \tau} b^\dagger b^\dagger bb \, e^{-\frac{i}{\hbar} \left( \hbar Q_L a^\dagger a \right) \tau} \ .
\end{aligned}
\tag{S67}
$$

We will use the identity:

$$be^{-\frac{i}{\hbar}(\hbar Q_L a^\dagger a)\tau} = (a-\lambda)e^{-\frac{i}{\hbar}(\hbar Q_L a^\dagger a)\tau} = ae^{-\frac{i}{\hbar}(\hbar Q_L a^\dagger a)\tau} - \lambda e^{-\frac{i}{\hbar}(\hbar Q_L a^\dagger a)\tau}$$

$$= e^{-iQ_L\tau}e^{-\frac{i}{\hbar}(\hbar Q_L a^\dagger a)\tau}a - e^{-\frac{i}{\hbar}(\hbar Q_L a^\dagger a)\tau}\lambda = e^{-iQ_L\tau}e^{-\frac{i}{\hbar}(\hbar Q_L a^\dagger a)\tau}(b+\lambda) - e^{-\frac{i}{\hbar}(\hbar Q_L a^\dagger a)\tau}\lambda$$

$$= e^{-iQ_L\tau}e^{-\frac{i}{\hbar}(\hbar Q_L a^\dagger a)\tau}b + e^{-iQ_L\tau}e^{-\frac{i}{\hbar}(\hbar Q_L a^\dagger a)\tau}\lambda - e^{-\frac{i}{\hbar}(\hbar Q_L a^\dagger a)\tau}\lambda$$

$$= e^{-iQ_L\tau}e^{-\frac{i}{\hbar}(\hbar Q_L a^\dagger a)\tau}b + (e^{-iQ_L\tau}-1)e^{-\frac{i}{\hbar}(\hbar Q_L a^\dagger a)\tau}\lambda \ . \tag{S68}$$

And similarly

$$b^\dagger e^{-\frac{i}{\hbar}(\hbar Q_L a^\dagger a)\tau} = (a^\dagger-\lambda)e^{-\frac{i}{\hbar}(\hbar Q_L a^\dagger a)\tau} = a^\dagger e^{-\frac{i}{\hbar}(\hbar Q_L a^\dagger a)\tau} - \lambda e^{-\frac{i}{\hbar}(\hbar Q_L a^\dagger a)\tau}$$

$$= e^{iQ_L\tau}e^{-\frac{i}{\hbar}(\hbar Q_L a^\dagger a)\tau}a^\dagger - e^{-\frac{i}{\hbar}(\hbar Q_L a^\dagger a)\tau}\lambda = e^{iQ_L\tau}e^{-\frac{i}{\hbar}(\hbar Q_L a^\dagger a)\tau}(b^\dagger+\lambda) - e^{-\frac{i}{\hbar}(\hbar Q_L a^\dagger a)\tau}\lambda$$

$$= e^{iQ_L\tau}e^{-\frac{i}{\hbar}(\hbar Q_L a^\dagger a)\tau}b^\dagger + e^{iQ_L\tau}e^{-\frac{i}{\hbar}(\hbar Q_L a^\dagger a)\tau}\lambda - e^{-\frac{i}{\hbar}(\hbar Q_L a^\dagger a)\tau}\lambda$$

$$= e^{iQ_L\tau}e^{-\frac{i}{\hbar}(\hbar Q_L a^\dagger a)\tau}b^\dagger + (e^{iQ_L\tau}-1)e^{-\frac{i}{\hbar}(\hbar Q_L a^\dagger a)\tau}\lambda \ . \tag{S69}$$

We have

$$\tilde{H}_I(\tau) = \hbar K_{\text{err}}e^{-iQ_L\tau}e^{\frac{i}{\hbar}(\hbar Q_L a^\dagger a)\tau}b^\dagger b^\dagger b e^{-\frac{i}{\hbar}(\hbar Q_L a^\dagger a)\tau}b + \lambda\hbar K_{\text{err}}(e^{-iQ_L\tau}-1)e^{\frac{i}{\hbar}(\hbar Q_L a^\dagger a)\tau}b^\dagger b^\dagger b e^{-\frac{i}{\hbar}(\hbar Q_L a^\dagger a)\tau} \ . \tag{S70}$$

Finally, we get

$$\begin{aligned}
\tilde{H}_I(\tau) = {}& \hbar K_{\text{err}}b^\dagger b^\dagger bb \\
&+ \lambda\hbar K_{\text{err}}(e^{iQ_L\tau}-1)e^{-iQ_L\tau}b^\dagger bb \\
&+ \lambda\hbar K_{\text{err}}(e^{iQ_L\tau}-1)e^{-2iQ_L\tau}e^{\frac{i}{\hbar}(\hbar Q_L a^\dagger a)\tau}b^\dagger e^{-\frac{i}{\hbar}(\hbar Q_L a^\dagger a)\tau}bb \\
&+ \lambda\hbar K_{\text{err}}(e^{-iQ_L\tau}-1)e^{-iQ_L\tau}e^{\frac{i}{\hbar}(\hbar Q_L a^\dagger a)\tau}b^\dagger b^\dagger e^{-\frac{i}{\hbar}(\hbar Q_L a^\dagger a)\tau}b \\
&+ \lambda\hbar K_{\text{err}}(e^{-iQ_L\tau}-1)e^{\frac{i}{\hbar}(\hbar Q_L a^\dagger a)\tau}b^\dagger b^\dagger b e^{-\frac{i}{\hbar}(\hbar Q_L a^\dagger a)\tau} \ .
\end{aligned} \tag{S71}$$

Moreover, similar techniques could be used to deal with terms with $\lambda$. We have

$$\begin{aligned}
&\lambda\hbar K_{\text{err}}\left(e^{iQ_L\tau}-1\right)e^{-2iQ_L\tau}e^{\frac{i}{\hbar}(\hbar Q_L a^\dagger a)\tau}b^\dagger e^{-\frac{i}{\hbar}(\hbar Q_L a^\dagger a)\tau}bb \\
&= \lambda\hbar K_{\text{err}}\left(e^{iQ_L\tau}-1\right)e^{-iQ_L\tau}b^\dagger bb + \lambda^2\hbar K_{\text{err}}\left(e^{iQ_L\tau}-1\right)^2 e^{-2iQ_L\tau}bb \ ,
\end{aligned} \tag{S72}$$

and

$$\begin{aligned}
&\lambda\hbar K_{\text{err}}\left(e^{-iQ_L\tau}-1\right)e^{-iQ_L\tau}e^{\frac{i}{\hbar}(\hbar Q_L a^\dagger a)\tau}b^\dagger b^\dagger e^{-\frac{i}{\hbar}(\hbar Q_L a^\dagger a)\tau}b \\
&= \lambda\hbar K_{\text{err}}\left(e^{-iQ_L\tau}-1\right)e^{\frac{i}{\hbar}(\hbar Q_L a^\dagger a)\tau}b^\dagger e^{-\frac{i}{\hbar}(\hbar Q_L a^\dagger a)\tau}b^\dagger b \\
&+ \lambda^2\hbar K_{\text{err}}\left(e^{-iQ_L\tau}-1\right)e^{-iQ_L\tau}\left(e^{iQ_L\tau}-1\right)e^{\frac{i}{\hbar}(\hbar Q_L a^\dagger a)\tau}b^\dagger e^{-\frac{i}{\hbar}(\hbar Q_L a^\dagger a)\tau}b \\
&= \lambda\hbar K_{\text{err}}\left(e^{-iQ_L\tau}-1\right)e^{iQ_L\tau}b^\dagger b^\dagger b \\
&+ 2\lambda^2\hbar K_{\text{err}}\left(e^{-iQ_L\tau}-1\right)\left(e^{iQ_L\tau}-1\right)b^\dagger b \\
&+ \lambda^3\hbar K_{\text{err}}\left(e^{-iQ_L\tau}-1\right)e^{-iQ_L\tau}\left(e^{iQ_L\tau}-1\right)^2 b \ ,
\end{aligned} \tag{S73}$$

and

$$\begin{aligned}
&\lambda\hbar K_{\text{err}}\left(e^{-iQ_L\tau}-1\right)e^{\frac{i}{\hbar}(\hbar Q_L a^\dagger a)\tau}b^\dagger b^\dagger b e^{-\frac{i}{\hbar}(\hbar Q_L a^\dagger a)\tau} \\
&= \lambda\hbar K_{\text{err}}\left(e^{-iQ_L\tau}-1\right)e^{iQ_L\tau}b^\dagger b^\dagger b \\
&+ \lambda^2\hbar K_{\text{err}}\left(e^{-iQ_L\tau}-1\right)\left(e^{iQ_L\tau}-1\right)b^\dagger b + \lambda^2\hbar K_{\text{err}}\left(e^{-iQ_L\tau}-1\right)\left(e^{iQ_L\tau}-1\right)b^\dagger b \\
&+ \lambda^2\hbar K_{\text{err}}\left(e^{-iQ_L\tau}-1\right)^2 e^{2iQ_L\tau}b^\dagger b^\dagger \\
&+ \lambda^3\hbar K_{\text{err}}\left(e^{-iQ_L\tau}-1\right)\left(e^{iQ_L\tau}-1\right)^2 e^{-iQ_L\tau}b \\
&+ 2\lambda^3\hbar K_{\text{err}}\left(e^{-iQ_L\tau}-1\right)^2\left(e^{iQ_L\tau}-1\right)e^{iQ_L\tau}b^\dagger \\
&+ \lambda^4\hbar K_{\text{err}}\left(e^{-iQ_L\tau}-1\right)^2\left(e^{iQ_L\tau}-1\right)^2 \ .
\end{aligned} \tag{S74}$$

Thus, collecting all the terms, we have

$$
\begin{aligned}
\mathcal{O}(\lambda^0) &: \hbar K_{\mathrm{err}} b^\dagger b^\dagger b b \,, \\
\mathcal{O}(\lambda^1) &: 2\lambda\hbar K_{\mathrm{err}}\left(1 - e^{-iQ_L\tau}\right) b^\dagger b b + 2\lambda\hbar K_{\mathrm{err}}\left(1 - e^{iQ_L\tau}\right) b^\dagger b^\dagger b \,, \\
\mathcal{O}(\lambda^2) &: \lambda^2\hbar K_{\mathrm{err}}\left(e^{iQ_L\tau} - 1\right)^2 e^{-2iQ_L\tau} b b + \lambda^2\hbar K_{\mathrm{err}}\left(e^{-iQ_L\tau} - 1\right)^2 e^{2iQ_L\tau} b^\dagger b^\dagger \\
&\quad + 4\lambda^2\hbar K_{\mathrm{err}}\left(e^{-iQ_L\tau} - 1\right)\left(e^{iQ_L\tau} - 1\right) b^\dagger b \,, \\
\mathcal{O}(\lambda^3) &: 2\lambda^3\hbar K_{\mathrm{err}}\left(e^{-iQ_L\tau} - 1\right)\left(e^{iQ_L\tau} - 1\right)^2 e^{-iQ_L\tau} b \\
&\quad + 2\lambda^3\hbar K_{\mathrm{err}}\left(e^{-iQ_L\tau} - 1\right)^2\left(e^{iQ_L\tau} - 1\right) e^{iQ_L\tau} b^\dagger \,, \\
\mathcal{O}(\lambda^4) &: \lambda^4\hbar K_{\mathrm{err}}\left(e^{-iQ_L\tau} - 1\right)^2\left(e^{iQ_L\tau} - 1\right)^2 \,.
\end{aligned}
\tag{S75}
$$

After reorganizing terms and performing integrals, we get

$$
-\frac{i}{\hbar}\int_0^T \tilde{H}_I(\tau)d\tau \left|\psi\right\rangle = \sum_\alpha g_0(\alpha)|\alpha\rangle_b + \sum_\alpha g_1(\alpha)b^\dagger|\alpha\rangle_b + \sum_\alpha g_2(\alpha)b^\dagger b^\dagger|\alpha\rangle_b \,,
\tag{S76}
$$

where

$$
\begin{aligned}
g_0(\alpha) &= K_{\mathrm{err}}\frac{\lambda^2 e^{-2iTQ_L}}{2Q_L}\left(\begin{array}{l}(\alpha+\lambda)^2 + e^{2iTQ_L}\left(3\alpha(\alpha+2) - 2iT\left(\alpha^2 + 6\alpha\lambda + 6\lambda^2\right)Q_L\right) \\ +4\lambda(\alpha+2\lambda)e^{3iTQ_L} - 4(\alpha+\lambda)(\alpha+2\lambda)e^{iTQ_L} + \lambda^2\left(-e^{4iTQ_L}\right)\end{array}\right) f(\alpha) \,, \\
g_1(\alpha) &= K_{\mathrm{err}}\frac{\lambda e^{-iTQ_L}}{Q_L}\left(\begin{array}{l}-2(\alpha+\lambda)^2 + e^{iTQ_L}\left(2\alpha^2 - 3\lambda^2 - 2iT(\alpha+\lambda)(\alpha+3\lambda)Q_L\right) \\ +2\lambda(2\alpha+3\lambda)e^{2iTQ_L} + \lambda^2\left(-e^{3iTQ_L}\right)\end{array}\right) f(\alpha) \,, \\
g_2(\alpha) &= \frac{K_{\mathrm{err}}}{2Q_L}\left(\lambda\left(-1 + e^{iTQ_L}\right)\left(4\alpha - \lambda\left(-3 + e^{iTQ_L}\right)\right) - 2iT(\alpha+\lambda)^2 Q_L\right) f(\alpha) \,.
\end{aligned}
\tag{S77}
$$

Moreover, we know that

$$
\begin{aligned}
|\delta\psi(\mathbf{x})\rangle &= e^{iT\left(\frac{\Omega^2}{\omega_m-\omega_L} - \frac{1}{2}\omega_m\right)} e^{-iQ_L T a^\dagger a}\sum_\alpha g_0(\alpha)|\alpha\rangle_b + e^{iT\left(\frac{\Omega^2}{\omega_m-\omega_L} - \frac{1}{2}\omega_m\right)} e^{-iQ_L T a^\dagger a}\sum_\alpha g_1(\alpha)b^\dagger|\alpha\rangle_b \\
&\quad + e^{iT\left(\frac{\Omega^2}{\omega_m-\omega_L} - \frac{1}{2}\omega_m\right)} e^{-iQ_L T a^\dagger a}\sum_\alpha g_2(\alpha)b^\dagger b^\dagger|\alpha\rangle_b \,.
\end{aligned}
\tag{S78}
$$

Since

$$
\begin{aligned}
&e^{iT\left(\frac{\Omega^2}{\omega_m-\omega_L} - \frac{1}{2}\omega_m\right)} e^{-iQ_L T a^\dagger a}\sum_\alpha g_0(\alpha)|\alpha\rangle_b \\
&= e^{iT\left(\frac{\Omega^2}{\omega_m-\omega_L} - \frac{1}{2}\omega_m\right) + i\lambda\mathrm{Im}\left[\lambda\left(1 - e^{iQ_L T}\right)\right]}\sum_\beta g_0\left((\beta+\lambda)e^{iQ_L T} - \lambda\right) e^{i\lambda\mathrm{Im}\left[\beta\left(1 - e^{iQ_L T}\right)\right]}|\beta\rangle_b \,.
\end{aligned}
\tag{S79}
$$

Furthermore, using the identity,

$$
e^{-iQ_L T a^\dagger a}b^\dagger = e^{-iQ_L T}b^\dagger e^{-iQ_L T a^\dagger a} + \left(e^{-iQ_L T} - 1\right) e^{-iQ_L T a^\dagger a}\lambda \,,
\tag{S80}
$$

we have

$$
\begin{aligned}
&e^{iT\left(\frac{\Omega^2}{\omega_m-\omega_L} - \frac{1}{2}\omega_m\right)} e^{-iQ_L a^\dagger a T}\sum_\alpha g_1(\alpha)b^\dagger|\alpha\rangle_b \\
&= e^{iT\left(\frac{\Omega^2}{\omega_m-\omega_L} - \frac{1}{2}\omega_m\right)} e^{-iQ_L T}\sum_\alpha g_1(\alpha)b^\dagger e^{-iQ_L T a^\dagger a}|\alpha\rangle_b \\
&\quad + \lambda e^{iT\left(\frac{\Omega^2}{\omega_m-\omega_L} - \frac{1}{2}\omega_m\right)}\left(e^{-iQ_L T} - 1\right)\sum_\alpha g_1(\alpha)e^{-iQ_L T a^\dagger a}|\alpha\rangle_b \\
&= e^{iT\left(\frac{\Omega^2}{\omega_m-\omega_L} - \frac{1}{2}\omega_m\right) + i\lambda\mathrm{Im}\left[\lambda\left(1 - e^{iQ_L T}\right)\right]} e^{-iQ_L T}\sum_\beta g_1\left((\beta+\lambda)e^{iQ_L T} - \lambda\right) e^{i\lambda\mathrm{Im}\left[\beta\left(1 - e^{iQ_L T}\right)\right]}b^\dagger|\beta\rangle_b \\
&\quad + \lambda e^{iT\left(\frac{\Omega^2}{\omega_m-\omega_L} - \frac{1}{2}\omega_m\right) + i\lambda\mathrm{Im}\left[\lambda\left(1 - e^{iQ_L T}\right)\right]}\left(e^{-iQ_L T} - 1\right)\sum_\beta g_1\left((\beta+\lambda)e^{iQ_L T} - \lambda\right) e^{i\lambda\mathrm{Im}\left[\beta\left(1 - e^{iQ_L T}\right)\right]}|\beta\rangle_b \,,
\end{aligned}
\tag{S81}
$$

and we have

$$
e^{-iQ_L T a^\dagger a} b^\dagger b^\dagger = e^{-iQ_L T} b^\dagger e^{-iQ_L T a^\dagger a} b^\dagger + \lambda \left( e^{-iQ_L T} - 1 \right) e^{-iQ_L T a^\dagger a} b^\dagger
$$

$$
= e^{-iQ_L T} b^\dagger \left( e^{-iQ_L T} b^\dagger e^{-iQ_L T a^\dagger a} + \left( e^{-iQ_L T} - 1 \right) e^{-iQ_L T a^\dagger a} \lambda \right)
$$

$$
+ \lambda \left( e^{-iQ_L T} - 1 \right) \left( e^{-iQ_L T} b^\dagger e^{-iQ_L T a^\dagger a} + \left( e^{-iQ_L T} - 1 \right) e^{-iQ_L T a^\dagger a} \lambda \right)
$$

$$
= e^{-2iQ_L T} b^\dagger b^\dagger e^{-iQ_L T a^\dagger a} + 2\lambda \left( e^{-2iQ_L T} - e^{-iQ_L T} \right) b^\dagger e^{-iQ_L T a^\dagger a}
$$

$$
+ \lambda^2 \left( e^{-2iQ_L T} - 2e^{-iQ_L T} + 1 \right) e^{-iQ_L T a^\dagger a} \; . \tag{S82}
$$

So we have

$$
e^{iT\left(\frac{\Omega^2}{\omega_m - \omega_L} - \frac{1}{2}\omega_m\right)} e^{-iQ_L T a^\dagger a} \sum_\alpha g_2(\alpha) b^\dagger b^\dagger |\alpha\rangle_b
$$

$$
= e^{iT\left(\frac{\Omega^2}{\omega_m - \omega_L} - \frac{1}{2}\omega_m\right)} e^{-2iQ_L T} \sum_\alpha g_2(\alpha) b^\dagger b^\dagger e^{-iQ_L T a^\dagger a} |\alpha\rangle_b
$$

$$
+ e^{iT\left(\frac{\Omega^2}{\omega_m - \omega_L} - \frac{1}{2}\omega_m\right)} 2\lambda \left( e^{-2iQ_L T} - e^{-iQ_L T} \right) \sum_\alpha g_2(\alpha) b^\dagger e^{-iQ_L T a^\dagger a} |\alpha\rangle_b
$$

$$
+ e^{iT\left(\frac{\Omega^2}{\omega_m - \omega_L} - \frac{1}{2}\omega_m\right)} \lambda^2 \left( e^{-2iQ_L T} - 2e^{-iQ_L T} + 1 \right) \sum_\alpha g_2(\alpha) e^{-iQ_L T a^\dagger a} |\alpha\rangle_b
$$

$$
= e^{iT\left(\frac{\Omega^2}{\omega_m - \omega_L} - \frac{1}{2}\omega_m\right) + i\lambda \mathrm{Im}\left[\lambda\left(1 - e^{iQ_L T}\right)\right]} e^{-2iQ_L T}
$$

$$
\sum_\beta g_2\left((\beta + \lambda) e^{iQ_L T} - \lambda\right) e^{i\lambda \mathrm{Im}\left[\beta\left(1 - e^{iQ_L T}\right)\right]} b^\dagger b^\dagger |\beta\rangle_b
$$

$$
+ e^{iT\left(\frac{\Omega^2}{\omega_m - \omega_L} - \frac{1}{2}\omega_m\right) + i\lambda \mathrm{Im}\left[\lambda\left(1 - e^{iQ_L T}\right)\right]} 2\lambda \left( e^{-2iQ_L T} - e^{-iQ_L T} \right)
$$

$$
\sum_\beta g_2\left((\beta + \lambda) e^{iQ_L T} - \lambda\right) e^{i\lambda \mathrm{Im}\left[\beta\left(1 - e^{iQ_L T}\right)\right]} b^\dagger |\beta\rangle_b
$$

$$
+ e^{iT\left(\frac{\Omega^2}{\omega_m - \omega_L} - \frac{1}{2}\omega_m\right) + i\lambda \mathrm{Im}\left[\lambda\left(1 - e^{iQ_L T}\right)\right]} \lambda^2 \left( e^{-2iQ_L T} - 2e^{-iQ_L T} + 1 \right)
$$

$$
\sum_\beta g_2\left((\beta + \lambda) e^{iQ_L T} - \lambda\right) e^{i\lambda \mathrm{Im}\left[\beta\left(1 - e^{iQ_L T}\right)\right]} |\beta\rangle_b \; . \tag{S83}
$$

Thus, we could define

$$
|\delta\psi(\mathbf{x})\rangle = \sum_\beta h_0(\beta) |\beta\rangle_b + \sum_\beta h_1(\beta) b^\dagger |\beta\rangle_b + \sum_\beta h_2(\beta) b^\dagger b^\dagger |\beta\rangle_b \; , \tag{S84}
$$

where

$$
h_0(\beta) = e^{iT\left(\frac{\Omega^2}{\omega_m - \omega_L} - \frac{1}{2}\omega_m\right) + i\lambda \mathrm{Im}\left[\lambda\left(1 - e^{iQ_L T}\right)\right]} e^{i\lambda \mathrm{Im}\left[\beta\left(1 - e^{iQ_L T}\right)\right]} g_0\left((\beta + \lambda) e^{iQ_L T} - \lambda\right)
$$

$$
+ \lambda e^{iT\left(\frac{\Omega^2}{\omega_m - \omega_L} - \frac{1}{2}\omega_m\right) + i\lambda \mathrm{Im}\left[\lambda\left(1 - e^{iQ_L T}\right)\right]} e^{i\lambda \mathrm{Im}\left[\beta\left(1 - e^{iQ_L T}\right)\right]} \left( e^{-iQ_L T} - 1 \right) g_1\left((\beta + \lambda) e^{iQ_L T} - \lambda\right)
$$

$$
+ \lambda^2 e^{iT\left(\frac{\Omega^2}{\omega_m - \omega_L} - \frac{1}{2}\omega_m\right) + i\lambda \mathrm{Im}\left[\lambda\left(1 - e^{iQ_L T}\right)\right]} e^{i\lambda \mathrm{Im}\left[\beta\left(1 - e^{iQ_L T}\right)\right]} \left( e^{-2iQ_L T} - 2e^{-iQ_L T} + 1 \right) g_2\left((\beta + \lambda) e^{iQ_L T} - \lambda\right) \; ,
$$

$$
h_1(\beta) = e^{iT\left(\frac{\Omega^2}{\omega_m - \omega_L} - \frac{1}{2}\omega_m\right) + i\lambda \mathrm{Im}\left[\lambda\left(1 - e^{iQ_L T}\right)\right]} e^{i\lambda \mathrm{Im}\left[\beta\left(1 - e^{iQ_L T}\right)\right]} e^{-iQ_L T} g_1\left((\beta + \lambda) e^{iQ_L T} - \lambda\right)
$$

$$
+ 2\lambda e^{iT\left(\frac{\Omega^2}{\omega_m - \omega_L} - \frac{1}{2}\omega_m\right) + i\lambda \mathrm{Im}\left[\lambda\left(1 - e^{iQ_L T}\right)\right]} e^{i\lambda \mathrm{Im}\left[\beta\left(1 - e^{iQ_L T}\right)\right]} \left( e^{-2iQ_L T} - e^{-iQ_L T} \right) g_2\left((\beta + \lambda) e^{iQ_L T} - \lambda\right) \; ,
$$

$$
h_2(\beta) = e^{iT\left(\frac{\Omega^2}{\omega_m - \omega_L} - \frac{1}{2}\omega_m\right) + i\lambda \mathrm{Im}\left[\lambda\left(1 - e^{iQ_L T}\right)\right]} e^{i\lambda \mathrm{Im}\left[\beta\left(1 - e^{iQ_L T}\right)\right]} e^{-2iQ_L T} g_2\left((\beta + \lambda) e^{iQ_L T} - \lambda\right) \; . \tag{S85}
$$

Thus we have

$$
|\delta\phi(\mathbf{x})\rangle = e^{-i\omega_L T b^\dagger b} |\delta\psi(\mathbf{x})\rangle = \sum_\beta s_0(\beta) |\beta\rangle_b + \sum_\beta s_1(\beta) b^\dagger |\beta\rangle_b + \sum_\beta s_2(\beta) b^\dagger b^\dagger |\beta\rangle_b \; , \tag{S86}
$$

where

$$s_0(\beta) = h_0\left(\beta e^{i\omega_L T}\right) \ ,$$
$$s_1(\beta) = h_1\left(\beta e^{i\omega_L T}\right) e^{-iT\omega_L} \ ,$$
$$s_2(\beta) = h_2\left(\beta e^{i\omega_L T}\right) e^{-2iT\omega_L} \ . \tag{S87}$$

Now, consider the kernel function,

$$\begin{aligned}
K\left(\mathbf{x}, \mathbf{x}'\right) &= \left|\langle\phi(\mathbf{x})|\phi\left(\mathbf{x}'\right)\rangle\right|^2 \\
&= \left|\left(\langle\phi_0(\mathbf{x})| + \langle\delta\phi(\mathbf{x})|\right)\left(|\phi_0\left(\mathbf{x}'\right)\rangle + |\delta\phi\left(\mathbf{x}'\right)\rangle\right)\right|^2 \\
&= \left|\langle\phi_0(\mathbf{x})|\phi_0\left(\mathbf{x}'\right)\rangle + \langle\delta\phi(\mathbf{x})|\phi_0\left(\mathbf{x}'\right)\rangle + \langle\phi_0(\mathbf{x})|\delta\phi\left(\mathbf{x}'\right)\rangle + \langle\delta\phi(\mathbf{x})|\delta\phi\left(\mathbf{x}'\right)\rangle\right|^2 \\
&= \left(\langle\phi_0(\mathbf{x})|\phi_0\left(\mathbf{x}'\right)\rangle + \langle\delta\phi(\mathbf{x})|\phi_0\left(\mathbf{x}'\right)\rangle + \langle\phi_0(\mathbf{x})|\delta\phi\left(\mathbf{x}'\right)\rangle + \langle\delta\phi(\mathbf{x})|\delta\phi\left(\mathbf{x}'\right)\rangle\right) \\
&\quad \left(\langle\phi_0\left(\mathbf{x}'\right)|\phi_0(\mathbf{x})\rangle + \langle\phi_0\left(\mathbf{x}'\right)|\delta\phi(\mathbf{x})\rangle + \langle\delta\phi\left(\mathbf{x}'\right)|\phi_0(\mathbf{x})\rangle + \langle\delta\phi\left(\mathbf{x}'\right)|\delta\phi(\mathbf{x})\rangle\right) \\
&= \left|\langle\phi_0(\mathbf{x})|\phi_0\left(\mathbf{x}'\right)\rangle\right|^2 + \langle\delta\phi(\mathbf{x})|\phi_0\left(\mathbf{x}'\right)\rangle\langle\phi_0\left(\mathbf{x}'\right)|\phi_0(\mathbf{x})\rangle \\
&\quad + \langle\phi_0(\mathbf{x})|\delta\phi\left(\mathbf{x}'\right)\rangle\langle\phi_0\left(\mathbf{x}'\right)|\phi_0(\mathbf{x})\rangle + \langle\phi_0(\mathbf{x})|\phi_0\left(\mathbf{x}'\right)\rangle\langle\phi_0\left(\mathbf{x}'\right)|\delta\phi(\mathbf{x})\rangle \\
&\quad + \langle\phi_0(\mathbf{x})|\phi_0\left(\mathbf{x}'\right)\rangle\langle\delta\phi\left(\mathbf{x}'\right)|\phi_0(\mathbf{x})\rangle + \mathcal{O}(K_{\mathrm{err}}^2) \\
&= \kappa_0(\mathbf{x}, \mathbf{x}') + \langle\delta\phi(\mathbf{x})|\phi_0\left(\mathbf{x}'\right)\rangle\langle\phi_0\left(\mathbf{x}'\right)|\phi_0(\mathbf{x})\rangle \\
&\quad + \langle\phi_0(\mathbf{x})|\delta\phi\left(\mathbf{x}'\right)\rangle\langle\phi_0\left(\mathbf{x}'\right)|\phi_0(\mathbf{x})\rangle + \langle\phi_0(\mathbf{x})|\phi_0\left(\mathbf{x}'\right)\rangle\langle\phi_0\left(\mathbf{x}'\right)|\delta\phi(\mathbf{x})\rangle \\
&\quad + \langle\phi_0(\mathbf{x})|\phi_0\left(\mathbf{x}'\right)\rangle\langle\delta\phi\left(\mathbf{x}'\right)|\phi_0(\mathbf{x})\rangle + \mathcal{O}(K_{\mathrm{err}}^2) \ . \tag{S88}
\end{aligned}$$

We only need to compute a single term, and other terms would follow the symmetric and hermitian operations. We know from the previous analysis

$$|\phi_0(\mathbf{x})\rangle = e^{-\frac{1}{2}i\omega_m T + iQ_L\lambda^2 T - i\lambda^2 \sin(Q_L T)} \sum_\beta f\left(\left(e^{i\omega_L T}\beta + \lambda\right)e^{iQ_L T} - \lambda\right)e^{i\lambda \mathrm{Im}\left[\beta e^{i\omega_L T}\left(1 - e^{iQ_L T}\right)\right]}|\beta\rangle_b \ . \tag{S89}$$

Finally we have

$$\langle\phi_0(\bar{\mathbf{x}})|\delta\phi(\mathbf{x})\rangle = \sum_{\beta,\bar{\beta}} G(\beta, \bar{\beta}; \mathbf{x}, \bar{\mathbf{x}})e_b(\beta, \bar{\beta}) \ , \tag{S90}$$

where

$$e_b(\beta_1, \beta_2) = \langle\beta_1|\beta_2\rangle_b = e^{\beta_1^* \beta_2 - |\beta_1|^2/2 - |\beta_2|^2/2} \ , \tag{S91}$$

and

$$\begin{aligned}
G(\beta, \bar{\beta}; \mathbf{x}, \bar{\mathbf{x}}) &= e^{-i\bar{T}\left(\frac{\Omega^2}{\omega_m - \bar{\omega}_L} - \frac{1}{2}\omega_m\right)} e^{i\frac{\bar{\Omega}^2}{(\omega_m - \bar{\omega}_L)^2}\sin\left((\omega_m - \bar{\omega}_L)\bar{T}\right)} e^{-i\frac{\bar{\Omega}}{(\omega_m - \bar{\omega}_L)}\mathrm{Im}\left[\bar{\beta}e^{i\bar{\omega}_L \bar{T}}\left(1 - e^{i(\omega_m - \bar{\omega}_L)\bar{T}}\right)\right]} \\
&\quad f^*\left(e^{i\omega_m \bar{T}}\bar{\beta} - \frac{\bar{\Omega}}{(\omega_m - \bar{\omega}_L)}\left(1 - e^{i(\omega_m - \bar{\omega}_L)\bar{T}}\right)\right)s_0(\mathbf{x}, \beta) \\
&\quad + \bar{\beta}e^{-i\bar{T}\left(\frac{\Omega^2}{\omega_m - \bar{\omega}_L} - \frac{1}{2}\omega_m\right)} e^{i\frac{\Omega^2}{(\omega_m - \bar{\omega}_L)^2}\sin\left((\omega_m - \bar{\omega}_L)\bar{T}\right)} e^{-i\frac{\Omega}{(\omega_m - \bar{\omega}_L)}\mathrm{Im}\left[\bar{\beta}e^{i\bar{\omega}_L \bar{T}}\left(1 - e^{i(\omega_m - \bar{\omega}_L)\bar{T}}\right)\right]} \\
&\quad f^*\left(e^{i\omega_m \bar{T}}\bar{\beta} - \frac{\bar{\Omega}}{(\omega_m - \bar{\omega}_L)}\left(1 - e^{i(\omega_m - \bar{\omega}_L)\bar{T}}\right)\right)s_1(\mathbf{x}, \beta) \\
&\quad + \bar{\beta}^2 e^{-i\bar{T}\left(\frac{\Omega^2}{\omega_m - \bar{\omega}_L} - \frac{1}{2}\omega_m\right)} e^{i\frac{\Omega^2}{(\omega_m - \bar{\omega}_L)^2}\sin\left((\omega_m - \bar{\omega}_L)\bar{T}\right)} e^{-i\frac{\Omega}{(\omega_m - \bar{\omega}_L)}\mathrm{Im}\left[\bar{\beta}e^{i\bar{\omega}_L \bar{T}}\left(1 - e^{i(\omega_m - \bar{\omega}_L)\bar{T}}\right)\right]} \\
&\quad f^*\left(e^{i\omega_m \bar{T}}\bar{\beta} - \frac{\bar{\Omega}}{(\omega_m - \bar{\omega}_L)}\left(1 - e^{i(\omega_m - \bar{\omega}_L)\bar{T}}\right)\right)s_2(\mathbf{x}, \beta) \ . \tag{S92}
\end{aligned}$$

Moreover, if we assume

$$\langle\phi_0(\bar{\mathbf{x}})|\phi_0(\mathbf{x})\rangle = G_0(\mathbf{x}, \bar{\mathbf{x}}) \ , \tag{S93}$$

we get

$$K\left(\mathbf{x}, \mathbf{x}'\right) = K_0\left(\mathbf{x}, \mathbf{x}'\right) + \delta K\left(\mathbf{x}, \mathbf{x}'\right) + \mathcal{O}\left(K_{\mathrm{err}}^2\right) \ , \tag{S94}$$

and,

$$\delta K\left(\mathbf{x}, \bar{\mathbf{x}}\right) = \left(\sum_{\beta, \bar{\beta}} G_0^*(\mathbf{x}, \bar{\mathbf{x}}) G(\beta, \bar{\beta}; \mathbf{x}, \bar{\mathbf{x}}) e_b(\beta, \bar{\beta})\right) + \left(\sum_{\beta, \bar{\beta}} G_0^*(\mathbf{x}, \bar{\mathbf{x}}) G^*(\bar{\beta}, \beta; \bar{\mathbf{x}}, \mathbf{x}) e_b(\beta, \bar{\beta})\right)$$
$$+ \left(\sum_{\beta, \bar{\beta}} G_0(\mathbf{x}, \bar{\mathbf{x}}) G^*(\beta, \bar{\beta}; \mathbf{x}, \bar{\mathbf{x}}) e_b(\beta, \bar{\beta})\right) + \left(\sum_{\beta, \bar{\beta}} G_0(\mathbf{x}, \bar{\mathbf{x}}) G(\bar{\beta}, \beta; \bar{\mathbf{x}}, \mathbf{x}) e_b(\beta, \bar{\beta})\right) . \tag{S95}$$

Moreover, we use the setup again where $\alpha = \alpha_i$. We have

$$|\phi_0(\mathbf{x})\rangle = e^{-\frac{1}{2} i \omega_m T + i \frac{\Omega^2}{Q_L} T - i \frac{\Omega(\alpha_i Q_L + \Omega)}{Q_L^2} \sin(Q_L T)} \left| e^{-i\omega_m T} \alpha_i + \left(1 - e^{iQ_L T}\right) \frac{\Omega}{Q_L} e^{-i\omega_m T} \right\rangle_b ,$$

$$\langle\phi_0(\bar{\mathbf{x}})| = e^{\frac{1}{2} i \omega_m \bar{T} - i \frac{\Omega^2}{Q_L} \bar{T} + i \frac{\Omega(\alpha_i Q_L + \Omega)}{Q_L^2} \sin(\bar{Q}_L \bar{T})} \left\langle e^{-i\omega_m \bar{T}} \alpha_i + \left(1 - e^{i\bar{Q}_L \bar{T}}\right) \frac{\bar{\Omega}}{\bar{Q}_L} e^{-i\omega_m \bar{T}} \right|_b . \tag{S96}$$

Thus,

$$K_0\left(\mathbf{x}, \bar{\mathbf{x}}\right) = |\langle\phi_0(\bar{\mathbf{x}})|\phi_0(\mathbf{x})\rangle|^2 = |e_b(\beta, \bar{\beta})|^2 , \tag{S97}$$

where

$$\beta = e^{-i\omega_m T} \alpha_i + \left(1 - e^{iQ_L T}\right) \frac{\Omega}{Q_L} e^{-i\omega_m T} ,$$

$$\bar{\beta} = e^{-i\omega_m \bar{T}} \alpha_i + \left(1 - e^{i\bar{Q}_L \bar{T}}\right) \frac{\bar{\Omega}}{\bar{Q}_L} e^{-i\omega_m \bar{T}} , \tag{S98}$$

and we define

$$p(\mathbf{x}) = e^{-\frac{1}{2} i \omega_m T + i \frac{\Omega^2}{\bar{Q}_L} T - i \frac{\Omega(\alpha_i Q_L + \Omega)}{Q_L^2} \sin(Q_L T)} ,$$

$$\langle\phi_0(\bar{\mathbf{x}})|\phi_0(\mathbf{x})\rangle = p^*(\bar{\mathbf{x}}) p(\mathbf{x}) e_b(\beta, \bar{\beta}) . \tag{S99}$$

Moreover, we have

$$\delta K\left(\mathbf{x}, \bar{\mathbf{x}}\right) = |e_b(\beta, \bar{\beta})|^2 \begin{pmatrix} p^*(\mathbf{x}) s_0(\mathbf{x}, \beta) \\ + p^*(\mathbf{x}) \bar{\beta} s_1(\mathbf{x}, \beta) \\ + p^*(\mathbf{x}) \bar{\beta}^2 s_2(\mathbf{x}, \beta) \\ + \text{c.c.} \\ + (\mathbf{x} \leftrightarrow \bar{\mathbf{x}}) \\ + (\mathbf{x} \leftrightarrow \bar{\mathbf{x}}, \text{c.c.}) \end{pmatrix} . \tag{S100}$$

There is a further simplification for the phase factor. We could define

$$\tilde{h}_0(\beta) = g_0\left((\beta + \lambda) e^{iQ_L T} - \lambda\right) + \lambda \left(e^{-iQ_L T} - 1\right) g_1\left((\beta + \lambda) e^{iQ_L T} - \lambda\right)$$
$$+ \lambda^2 \left(e^{-2iQ_L T} - 2 e^{-iQ_L T} + 1\right) g_2\left((\beta + \lambda) e^{iQ_L T} - \lambda\right) ,$$
$$\tilde{h}_1(\beta) = e^{-iQ_L T} g_1\left((\beta + \lambda) e^{iQ_L T} - \lambda\right) + 2\lambda \left(e^{-2iQ_L T} - e^{-iQ_L T}\right) g_2\left((\beta + \lambda) e^{iQ_L T} - \lambda\right) ,$$
$$\tilde{h}_2(\beta) = e^{-2iQ_L T} g_2\left((\beta + \lambda) e^{iQ_L T} - \lambda\right) , \tag{S101}$$

and

$$\tilde{s}_0(\beta) = \tilde{h}_0\left(\beta e^{i\omega_L T}\right) ,$$
$$\tilde{s}_1(\beta) = \tilde{h}_1\left(\beta e^{i\omega_L T}\right) e^{-iT\omega_L} ,$$
$$\tilde{s}_2(\beta) = \tilde{h}_2\left(\beta e^{i\omega_L T}\right) e^{-2iT\omega_L} , \tag{S102}$$

we have

$$\delta K(\mathbf{x}, \bar{\mathbf{x}}) = |e_b(\beta, \bar{\beta})|^2 \begin{pmatrix} \tilde{s}_0(\mathbf{x}, \beta) + \bar{\beta} \tilde{s}_1(\mathbf{x}, \beta) + \bar{\beta}^2 \tilde{s}_2(\mathbf{x}, \beta) \\ + \text{c.c.} \\ + (\mathbf{x} \leftrightarrow \bar{\mathbf{x}}) \\ + (\mathbf{x} \leftrightarrow \bar{\mathbf{x}}, \text{c.c.}) \end{pmatrix} . \tag{S103}$$

So,

$$\frac{\delta K(\mathbf{x}, \bar{\mathbf{x}})}{K_0(\mathbf{x}, \bar{\mathbf{x}})} = \begin{pmatrix} \tilde{s}_0(\mathbf{x}, \beta) + \bar{\beta}\tilde{s}_1(\mathbf{x}, \beta) + \bar{\beta}^2\tilde{s}_2(\mathbf{x}, \beta) \\ +\text{c.c.} \\ +(\mathbf{x} \leftrightarrow \bar{\mathbf{x}}) \\ +(\mathbf{x} \leftrightarrow \bar{\mathbf{x}}, \text{c.c.}) \end{pmatrix} . \tag{S104}$$

## IV.    EXPERIMENTAL PROPOSALS

### A.    Threshold of Kerr non-linearity

As stated in the main text, we compare the Kerr coefficient with the drive amplitude to determined if the system is in the weak or strong non-linear regime. The Kerr non-linearity effectively induces a photon number-dependent resonant frequency shift $\omega'_{\mathrm{m}} \to (\omega_{\mathrm{m}} - 2K_{\mathrm{err}}n)$ with $n$ being the photon number; namely, every time a photon is added into the resonator, the resonant frequency will shift by $-2K_{\mathrm{err}}$. If we consider an initially on-resonance drive with $\omega_{\mathrm{L}} = \omega_{\mathrm{m}}$, as more and more photons are populated in the resonator, the drive will equivalently become more and more detuned. Strong non-linearity refers to the situation that after the first photon enters the resonator, the drive detuning $\Delta_{\mathrm{L}} \approx 2K_{\mathrm{err}}$ becomes large enough so that additional photons at $\omega_{\mathrm{L}}$ can no longer populate the resonator. In other words, the anharmonic resonator now may be approximated as a two-level qubit with negligible excitations in its higher energy levels. The intra-cavity photon number ($n \equiv b^\dagger b$) under a detuned drive tone is given by

$$n = \frac{\Omega^2}{\Delta_{\mathrm{L}}^2 + (\frac{\kappa}{2})^2}, \tag{S105}$$

where $\kappa = \kappa_{\mathrm{i}} + \kappa_{\mathrm{e}}$ is the full linewidth of the resonator. In our theoretical analysis, we assumed the intrinsic loss of the resonator is negligible $\kappa_{\mathrm{i}} \approx 0$ and the external coupling $\kappa_{\mathrm{e}}$ can also be kept sufficiently small. Therefore, the strong non-linearity regime corresponds to $\Omega^2/4K_{\mathrm{err}}^2 \ll 1$, while the weak non-linearity regime corresponds to $\Omega^2/4K_{\mathrm{err}}^2 \gg 1$.

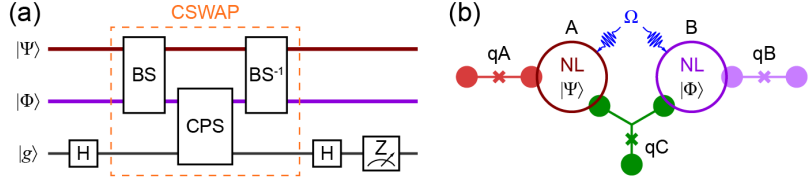

FIG. S1. (a) Measurement scheme for the overlap between two quantum state $|\Psi\rangle$ and $|\Phi\rangle$. (b) Experimental setup of a circuit QED system for the overlap measurement between the two modes supported by non-linear (NL) resonators A and B, respectively. Three transmon qubits qA, qB, and qC are used for quantum state control and coupling of the two resonators.

### B.    Experimental setup

The kernel we defined in Eq. (1) in the main text relies on the measurement of the overlap between two states $|\Psi\rangle \equiv |\psi(T)\rangle$ and $|\Phi\rangle \equiv |\psi(T')\rangle$. This can be realized by controlled-SWAP (CSWAP) gate using an auxiliary qubit [20]; the states of $|\Psi\rangle$ and $|\Phi\rangle$ are swapped when the qubit is in the excited ($|e\rangle$) state, while kept unchanged when the qubit is in the ground ($|g\rangle$) state. The CSWAP can be decomposed into two 50:50 beam-splitter (BS) operations with a controlled phase gate (CPS) in the middle (Fig. S1(a)). Therefore, if we first apply a Hadamard gate (H) to prepare the qubit to a superposition state $(|g\rangle + |e\rangle)/\sqrt{2}$, then apply the controlled SWAP gate, followed by a second Hadamard gate (H), the system state becomes

$$|g\rangle |\Psi\rangle |\Phi\rangle \xrightarrow{\mathrm{H}} \frac{\sqrt{2}}{2}(|g\rangle + |e\rangle) |\Psi\rangle |\Phi\rangle \xrightarrow{\mathrm{CSWAP}} \frac{\sqrt{2}}{2}(|g\rangle |\Psi\rangle |\Phi\rangle + |e\rangle |\Phi\rangle |\Psi\rangle)$$

$$\xrightarrow{\mathrm{H}} \frac{1}{2} |g\rangle (|\Psi\rangle |\Phi\rangle + |\Phi\rangle |\Psi\rangle) + \frac{1}{2} |e\rangle (|\Psi\rangle |\Phi\rangle - |\Phi\rangle |\Psi\rangle). \tag{S106}$$

Therefore, if we measure the probability of the qubit final state in $|g\rangle$ or $|e\rangle$, we can get the overlap $|\langle\Psi|\Phi\rangle|^2$

$$P_g = \frac{1}{2}(1 + |\langle\Psi|\Phi\rangle|^2), P_e = \frac{1}{2}(1 - |\langle\Psi|\Phi\rangle|^2). \tag{S107}$$

Experimentally, these quantum operations and measurements can be done in circuit QED systems [21, 22]. As illustrated in Fig. S1(b), we can use two nonlinear superconducting resonators (A and B) to prepare the states $|\Psi\rangle$ and $|\Phi\rangle$. The Kerr nonlinearity can be realized by either Josephson junction elements or superconducting kinetic inductance devices [23–25]. The drive tone $\Omega$ can be directly sent to the cavities via a microwave port. In addition, each cavity is dispersively coupled to a transmon qubit (qA and qB, respectively) for universal control and state readout. Importantly, a third transmon qC is simultaneously coupled to A and B to mediate a bilinear interaction $H_{BS} \propto (gab^\dagger + g^*a^\dagger b)$ under two microwave drive tones, which enables the beam-splitter interaction. The CPS gate can be realized by harnessing the dispersive coupling, for example, between qB and cavity B. Then we will be able to infer the state overlap by measuring the population of qB.

---


}

\end{document}